\documentclass[intlimits,twoside,a4paper]{article}

\usepackage{amsmath,amssymb}
\usepackage{graphicx}
\usepackage{wrapfig}
\usepackage{color}

\usepackage[T2A]{fontenc}
\usepackage[cp1251]{inputenc}

\usepackage[eqsecnum]{cmpj2}

\issue{2016}{19}{4}{43705}
\doinumber{10.5488/CMP.19.43705}

\title[BBGKY chain of  kinetic equations, non-equilibrium statistical operator method]
{BBGKY chain of  kinetic equations, non-equilibrium statistical operator method and
collective variable method in the statistical theory of non-equilibrium liquids}

\author[I.R.~Yukhnovskii, P.A.~Hlushak, M.V.~Tokarchuk]{I.R.~Yukhnovskii, P.A.~Hlushak,
 M.V.~Tokarchuk}

\address{Institute for Condensed Matter Physics
of the National Academy of Sciences of Ukraine,\\
1 Svientsitskii St., 79011 Lviv, Ukraine}

\authorcopyright{I.R.~Yukhnovskii, P.A.~Hlushak, M.V.~Tokarchuk, 2016}
\date{Received July 6, 2016, in final form November 1, 2016}

\begin{document}

\maketitle

\begin{abstract}
A chain of kinetic equations for non-equilibrium one-particle, two-particle and $ s $-particle
distribution functions of particles which take into account nonlinear hydrodynamic fluctuations
is proposed.
The method of Zubarev non-equilibrium statistical operator with projection is used.
Nonlinear hydrodynamic fluctuations are described with non-equilibrium distribution function
of collective variables that satisfies  generalized Fokker-Planck equation.
On the basis of the method of collective variables, a scheme of calculation of non-equilibrium
structural distribution function of collective variables and their hydrodynamic speeds
(above Gaussian approximation) contained in the generalized Fokker-Planck equation for the non-equilibrium distribution function of collective variables is proposed.
Contributions of short- and long-range interactions between particles are separated,
so that the short-range interactions (for example, the model of hard spheres) are described
in the coordinate space, while the long-range interactions --- in the space of collective variables.
Short-ranged component is regarded as basic, and corresponds to the
BBGKY chain of equations
for the model of hard spheres.

\keywords  non-linear fluctuations, non-equilibrium statistical operator,
distribution function, Fokker–Planck equation, simple fluid
 \pacs 74.40.Gh, 05.70.L, 64.70.F

\end{abstract}

\section{Introduction}

A development of equilibrium and non-equilibrium statistical mechanics of classical and quantum systems, which began in 1970-ies following the work by Bogoliubov \cite{Bogoliub1964} and
works by Born, Green \cite{BornGreen1949}, Kirkwood \cite{Kirkwood1946,Kirkwood1947},
Yvon \cite{Yvon1935} and continues today, has led to a substantial progress in the theory
of gases, liquids, plasma.
Bogoliubov in the book \cite{Bogoliub1964}, which turns 70 years, in a strict form formulated
the idea of a hierarchy of relaxation times in a system of many interacting particles and
a reduced number of parameters describing the evolution of the system.
This idea played a major role in the development of modern methods of non-equilibrium theory
to study the dynamics of macroscopic systems at kinetic and hydrodynamic stages, based on
the  fundamental principles of statistical mechanics.

Important for the development of this direction were the works by Zubarev
\cite{Zubarev1961,Zubarev1965,Zubarev1970}, Zwanzig \cite{Zwanzig1961,Zwanzig1963,Zwanzig1964},
Robertson \cite{Robertson1967-1,Robertson1967-2}, Kawasaki and Gunton
\cite{KawasGunton1973}, Peletminskii and Yatsenko \cite{PeletJats1967}, Zubarev and
Kalashnikov \cite{ZubarKalash1970}.
The results of the research
done in this field are detailed in books
\cite{Zubarev1971,AkhiezPelet1977,ZubMorRop-1,ZubMorRop-2,LapilKalash2008,
KostTokMarkIgnGnat2009,KhamzNigmat2011} and reviews
\cite{Zubarev1980,ZubarZerkov1986}.
However, along with the important results of  theories in statistical physics,
as well as in other fields of modern physics, there are still many unsolved problems,
especially in the theory of non-equilibrium processes.

 In dense gases and fluids in the field of phase transitions, the heterophase fluctuations play an
important role
\cite{BogShumYuk1984,ShumYuk1985,Yukalov1991,Fisher1993,OlemKopl1995,Olemskoi1999,BakFish2004,Onuki2004,Das2011}.
They always appear and disappear during the diffusion processes.
In the field of phase transitions, the heterophase fluctuations are factors forming a new phase,
in particular, forming the bubbles in a liquid or drops in a gas.

The non-equilibrium gas–liquid phase transition is characterized by nonlinear hydrodynamic
fluctuations of mass, momentum and particle energy, which describe a collective nature of the
process and define the spatial and temporal behavior of the transport coefficients (viscosity,
thermal conductivity), time correlation functions and a dynamic structure factor.
At the same time, due to heterogeneity in collective dynamics of these fluctuations,
liquid drops emerge in the gas phase (in case of transition from the gas phase to the liquid
phase), or the gas bubbles emerge in the liquid phase (in case of transition from
the liquid phase to the gas phase), the formation of which is of a kinetic nature described
by a redistribution of momentum and energy, i.e., when a certain group of particles
in the system receives a significant decrease (in the case of drops),
or increase (in the case of bubbles) of kinetic energy. The particles, that
form bubbles or droplets, diffuse out of their phases in the liquid or in the gas and vice versa.
They have different values of momentum, energy and pressure in different phases.
All these features are related to the non-equilibrium one-, two-,
$s$-particle distribution functions (which depend on the coordinate,
momentum and time) that satisfy the Bogoliubov-Born-Green-Kirkwood-Yvon (BBGKY)
chain of equations.
They are the real heterophase systems in which bubble embryoses, drops or small crystals
are of a kinetic nature caused by nonlinear fluctuations, changes in temperature, pressure.
In our case, the heterophase formations (containing a finite particle number in this or that
phase) can be described by non-equilibrium distribution function $f_{s}(x^{s};t)$.
The kinetic processes within heterophase formations are described respectively by the kinetic
equations \cite{HlushTokar2016}, in which the right-hand side contains the summands that take into account
the mutual effect of kinetic and hydrodynamic processes.
Obviously, such heterophase formations form and disappear (with finite lifetime),
exchanging  by both the particles and energy with  the surrounding particles  in the background
of nonlinear hydrodynamic fluctuations of densities of  particle number, momentum, energy;
the contribution of such fluctuations grows at phase transformations.
These nonlinear fluctuations are  described by the Fokker-Planck equation \cite{HlushTokar2016}.
Here, in the processes of interaction between kinetic and hydrodynamic fluctuations
with appropriate changes of temperature and pressure in heterophase formations due to
spontaneous symmetry breaking there can be self-organizing processes of sorts of particle motion
with  group velocity $f_{s}(x^{s};t)=f_{s}({\bf r}_{1}-{\bf v}t, {\bf p}_{1},\ldots,
{\bf r}_{s}-{\bf v}t, {\bf p}_{s})$, which leads to an automodel (quasi-soliton) spreading  of
heterophase formations in the respective system.

Such processes require a detailed separate study due to the calculation difficulties of kinetic
and hydrodynamic transport kernels in equations of transport. In this connection
we would like to draw attention to the Klimontovich article \cite{Klimontov1998},
in which to a certain extent there is realized a consistent description of kinetics
and hydrodynamics (diffusion processes are taken into account) for gas-liquid
phase transition.

The main difficulty is that the kinetics and hydrodynamics of these processes
are strongly interrelated and should be considered simultaneously.
In articles \cite{ZubMor1984,ZubMorOmTok1991,ZubMorOmTok1993} there is proposed a consistent description of kinetic and hydrodynamic
processes in dense gases and liquids on the basis of Zubarev non-equilibrium
statistical operator \cite{ZubMorRop-1,ZubMorRop-2}.

This article is an extension or generalization of article  \cite{HlushTokar2016}.
Here, we want to emphasize certain aspects.
In \cite{HlushTokar2016}, the main parameter for the description of kinetic processes was one-particle non-equi\-lib\-rium distribution function that corresponds to the gas description.
In order to get a consistent description of kinetic and hydrodynamic processes (which is important for dense gases and liquids \cite{ZubMor1984,ZubMorOmTok1991,ZubMorOmTok1993}) it is necessary to include the density of average potential energy as an important parameter.
Having included this parameter, we can identify explicit contributions connected with short-
and long-range interactions between particles.
For dilute gases, this parameter can be neglected, while for dense gases and liquids one makes
an important contribution compared to kinetic energy.
In this contribution, we will develop an approach \cite{HlushTokar2016} for a
consistent description of kinetic and hydrodynamic processes that are characterized
by non-linear fluctuations taking into account short-range and long-range interactions
between particles.
The approach is important for the description of non-equilibrium gas–liquid phase transition.

In Section~\ref{sec2}, we will obtain the non-equilibrium statistical operator for non-equilibrium
state of the system when the parameters of a reduced description are represented by
a non-equilibrium one-particle distribution function,
the density of non-equilibrium average value of
potential energy of particle interaction
and a distribution function of non-equilibrium
nonlinear hydrodynamic variables. Using this operator, we construct kinetic equations
for non-equilibrium one-, two-, s-particle distribution functions which
take into account nonlinear hydrodynamic fluctuations, for which the non-equilibrium
distribution function satisfies a generalized Fokker–Planck equation.

In Section~\ref{sec3}, we will consider a scheme for calculation of the structural distribution function
of hydrodynamic collective variables and their hydrodynamic velocities
(in approximation higher than Gaussian), which enter the generalized
Fokker–Planck equation for the non-equilibrium distribution function of hydrodynamic
collective variables.
We separate the contributions from short-range and long-range interactions between particles,
which will be described in the coordinate space and in the space of collective variables, respectively.
Moreover, the short-range component will be considered in a simplified manner,
which in our case will correspond to the BBGKY chain of equations for the model of
hard spheres \cite{KobrOmelTok1998}.

\section{Non-equilibrium distribution function and BBGKY chain of kinetic eq\-uations in the Zubarev non-equilibrium statistical operator method}
\label{sec2}

For a consistent description of kinetic and hydrodynamic fluctuations in a classical one-component
fluid, it is necessary to select the description parameters for one-particle and collective
processes.
For these parameters, we choose the non-equilibrium one-particle distribution function
$f_{1}(x;t)=\langle\hat{n}_{1}(x)\rangle^{t}$,
the density of non-equilibrium average value of potential energy of particle interaction
$H^{\text{int}}({\bf r},t)=$ \linebreak $\langle \hat{H}^{\text{int}}({\bf r})\rangle^{t}$
and distribution function of hydrodynamic variables
$f(a;t)=\langle\delta(\hat{a}-a)\rangle^{t}$. Here, the phase function
\begin{equation}    \label{eq2.1}
\hat{n}_{1}(x)=\sum_{j=1}^{N}\delta(x-x_{j})=
\sum_{j=1}^{N}\delta({\bf r}-{\bf r}_{j})\delta({\bf p}-{\bf p}_{j})
\end{equation}
is the microscopic particle number density.
$x_{j}=({\bf r}_{j},{\bf p}_{j})$ is the set of phase variables (coordinates and momenta),
$N$ is the total number of particles in a volume $V$. %
\begin{equation}    \label{eq2.2}
\hat{H}^{\text{int}}({\bf r})=\frac{1}{2}\sum_{j\neq l=1}^{N}\Phi(|{\bf r}_{l j}|)\delta({\bf r}-{\bf r}_{j})
\end{equation}
is the microscopic density of the potential energy  of particle interaction.
A microscopic phase distribution of hydrodynamic variables is given by
\begin{equation}    \label{eq2.3}
\hat{f}(a)=\delta (\hat{a}-a)=
\prod_{l=1}^{3}\prod_{\bf {k}}\delta (\hat{a}_{l \bf {k}}-a_{l \bf {k}}),
\end{equation}
where $\hat{a}_{1 \bf {k}}=\hat{n}_{\bf {k}}\,$, $\hat{a}_{2 \bf {k}}=\hat{\bf {J}}_{\bf {k}}\,$,
$\hat{a}_{3 \bf{k}}=\hat{\varepsilon}_{\bf {k}}$ are the Fourier components of the densities
of particle number, momentum and energy:
\begin{eqnarray}    \label{eq2.4}
\hat{n}_{\bf k}=\sum_{j=1}^{N}\re^{-\ri{\bf k} {\bf r}_{j}},  \qquad
\hat{\bf J}_{\bf k}=\sum_{j=1}^{N}{\bf p}_{j}\re^{-\ri{\bf k} {\bf r}_{j}},  \qquad
\hat{\varepsilon}_{\bf k}= \sum_{j=1}^{N}\left[ \frac{p^{2}_{j}}{2m}+\frac{1}{2}\sum_{l\neq
j=1}^{N}\Phi(|{\bf r}_{l j}|)\right]
 \re^{-\ri{\bf k} {\bf r}_{j}},
\end{eqnarray}
and $a_{m{\bf k}}=(n_{{\bf k}},{\bf J}_{\bf{k}},\varepsilon_{\bf k})$ are the corresponding
collective variables. $\Phi(|{\bf r}_{lj}|)=\Phi(|{\bf r}_{l}-{\bf r}_{j}|)$ is the pair
interaction potential between particles, which we assume can be represented as the sum
of short-range interaction $\Phi^{\text{sh}}(|{\bf r}_{lj}|)$ and long-range interaction
$\Phi^{\text{long}}(|{\bf r}_{lj}|)$ potentials:
\[
\Phi(|{\bf r}_{l j}|)=\Phi^{\text{sh}}(|{\bf r}_{l j}|)+\Phi^{\text{long}}(|{\bf r}_{l j}|).
\]
The introduction of $\langle\hat{H}^{\text{int}}({\bf r})\rangle^{t}$ as a parameter is important
step, as for liquids and dense gases it gives a larger contribution
than the kinetic part of energy connected with $\langle\hat{n}_{1}(x)\rangle^{t}$.
In addition this parameter describes the collective dynamics 
of short- and long-range interactions:
\[
\langle\hat{H}^{\text{int}}({\bf r})\rangle^{t}=\langle\hat{H}^{\text{sh}}({\bf r})\rangle^{t}+
\frac{1}{2V^2}\sum_{\bf q ,\bf k}\nu({\bf k})\re^{\ri {\bf qr}}
\left(\langle \hat{n}_{\bf q+\bf k} \hat{n}_{-\bf k}\rangle^{t}-
  \langle \hat{n}_{\bf q}\rangle^{t}\right),
\]
$\langle \hat{n}_{{\bf q}+{\bf k}} \hat{n}_{-\bf k}\rangle^{t}=F({\bf q},{\bf k};t)$ is the
non-equilibrium scattering function which connected with non-equilibrium dynamic factor of system,
$\nu({\bf k})=\int\rd{\bf r}\,\re^{-\ri{\bf k}{\bf r}}\Phi^{\text{long}}(|{\bf r}|)$ is the
Fourier component of long-range potential of particle interactions.
The average values $\langle\hat{n}_{1}(x)\rangle^{t}$,
$\langle \hat{H}^{\text{int}}({\bf r}) \rangle^t$ and
$\langle\delta(\hat{a}-a)\rangle^{t}$ are calculated by means of the non-equilibrium $N$-particle
distribution function $\varrho(x^{N};t)$, that satisfies the Liouville equation.
In line with the idea of reduced description of non-equilibrium states this function
is the functional
\begin{equation}    \label{Eq2.5}
\varrho \big(x^{N};t\big)=\varrho \big(\ldots,f_{1}(x;t),\langle \hat{H}^{\text{int}}({\bf r})\rangle^t,
                  f(a;t),\ldots\big).
\end{equation}
In order to find a non-equilibrium distribution function $\varrho(x^{N};t)$ we use
Zubarev's method \cite{ZubMorRop-1,ZubMorRop-2,ZubarMoroz1989},
in which a general solution of Liouville equation taking into account a projection procedure
can be presented in the form:
\begin{equation}    \label{Eq2.6}
\varrho\big(x^{N};t\big)=\varrho_{\text{rel}}\big(x^{N};t\big)-\int_{-\infty}^{t}\rd t'
  \re^{\epsilon(t'-t)}T(t,t')\big[1-P_{\text{rel}}(t')\big]\ri L_{N}\varrho_{\text{rel}}\big(x^{N};t'\big),
\end{equation}
where $\epsilon\rightarrow +0$ after thermodynamic limiting transition.
The source selects the retarded solutions of Liouville equation with operator $\ri L_{N}$.
$T(t,t')=\exp_{+}\{-\int_{t'}^{t}\rd t'[1-P_{\text{rel}}(t')]\ri L_{N}\}$ is the generalized time evolution
operator taking into account Kawasaki-Gunton projection $P_{\text{rel}}(t')$.
The structure of $P_{\text{rel}}(t')$ depends on the relevant distribution function
$\varrho_{\text{rel}}(x^{N};t)$, which in method by Zubarev is determined from extremum
of the information entropy at simultaneous conservation of normalization condition
\begin{equation}    \label{Eq2.7}
 \int \rd \Gamma_{N}\varrho_{\text{rel}} \big(x^{N};t\big)=1, \qquad \rd\Gamma_{N}=
 \frac{(\rd x)^{N}}{\mbox{h}^{3N}N!}=\frac{(\rd x_{1},\ldots,
 \rd x_{N})}{\mbox{h}^{3N}N!}\,, \qquad
  \rd x=\rd{\bf r} \rd{\bf p},
\end{equation}
and the fact that the parameters of the reduced description,
$f_{1}(x;t)$ and $f(a;t)$ are fixed.
Then, a relevant distribution function can be written as follows:
\begin{equation}    \label{Eq2.8}
 \varrho_{\text{rel}} \big(x^{N};t\big)=\exp \left[-\Phi(t)-\int \rd x \gamma(x;t)\hat{n}_{1}(x)-
 \int \rd{\bf r}\beta({\bf r};t) \hat{H}^{\text{int}}({\bf r})-
 \int \rd a F(a;t)\hat{f}(a)\right],
\end{equation}
where $\rd a$ is the integration over collective variables:
\begin{eqnarray}     \nonumber
\rd a = \prod\limits_{\bf k}\rd n_{\bf k}\rd{\bf j}_{\bf k}\rd\varepsilon_{\bf k}\,, \qquad
  \rd n_{\bf k}=\rd\Re n_{\bf k}\,\rd\Im n_{\bf k}\,,      \qquad
\rd \varepsilon_{\bf k} = \rd\Re \varepsilon_{\bf k}\,\rd\Im\varepsilon_{\bf k}\,,  \\  \nonumber
\rd{\bf j}_{\bf k} = \rd j_{x,{\bf k}}\,\rd j_{y,{\bf k}}\,\rd j_{z,{\bf k}}\,,
\qquad
\rd j_{x,{\bf k}}=\rd\Re j_{x,{\bf k}}\,\rd\Im j_{x,{\bf k}}\,, \,\,\,\ldots \,.
\end{eqnarray}
The Massieu-Planck functional $\Phi(t)$ is determined from the normalization condition
for the relevant distribution function
\[
 \Phi (t)=\ln \int \rd\Gamma_{N}\exp \left[-\int \rd x \gamma (x;t)\hat{n}_{1}(x)-
 \int \rd{\bf r}\beta({\bf r};t) \hat{H}^{\text{int}}({\bf r})-
 \int \rd a F(a;t)\hat{f}(a)\right].
\]

The functions $\gamma(x;t)$, $\beta({\bf r};t)$ and $F(a;t)$ are the Lagrange multipliers
and are determined from the self-consistent conditions:
\begin{eqnarray}    \label{Eq2.9}
 f_{1}(x;t)=\langle \hat{n}_{1}(x) \rangle^{t}=\langle \hat{n}_{1}(x) \rangle^{t}_{\text{rel}}\,, \quad
 \langle \hat{H}^{\text{int}}({\bf r})\rangle^{t}=\langle \hat{H}^{\text{int}}({\bf r})\rangle^{t}_{\text{rel}}\,,
\nonumber\\
  f(a;t)=\langle \delta (\hat{a}-a) \rangle^{t}=\langle \delta (\hat{a}-a) \rangle^{t}_{\text{rel}}\,,
                                                    \qquad \qquad
\end{eqnarray}
where $\langle \ldots \rangle^{t}=\int \rd\Gamma_{N} \ldots \varrho(x^{N};t)$ and
$\langle \ldots \rangle^{t}_{\text{rel}}=\int \rd\Gamma_{N} \ldots \varrho_{\text{rel}}(x^{N};t)$.
To find the explicit form of non-equi\-librium distribution function $\varrho(x^{N};t)$,
we exclude the factor $F(a;t)$ in a relevant distribution function and thereafter,
by means of self-consistent conditions (\ref{Eq2.9}), we have
\begin{equation}    \label{Eq2.10}
 \varrho_{\text{rel}} \big(x^{N};t\big)=\varrho_{\text{rel}}^{\text{kin-hyd}}\big(x^{N};t\big)
  \frac{f(a;t)}{W(a;t)}\Bigg{|}_{a=\hat{a}}\,.
\end{equation}
Here,
\begin{align}    \label{Eq2.11}
 W(a;t)&=\int \rd\Gamma_{N}\exp\left[-\Phi^{\text{kin-hyd}}(t)-
   \int \rd x \gamma(x;t)\hat{n}_{1}(x)-\int \rd{\bf r} \beta({\bf r};t)\hat{H}^{\text{int}}({\bf r})\right]
   \hat{f}(a) \nonumber\\
 &=\int \rd\Gamma_{N}\varrho_{\text{rel}}^{\text{kin-hyd}}\big(x^{N};t\big)\hat{f}(a)
\end{align}
is the structure distribution function of hydrodynamic variables, which
could be also considered as a Jacobian for transition from $\hat{f}(a)$ into
space of collective variables $n_{\bf{k}}$, $\bf{J}_{\bf{k}}$, $\varepsilon_{\bf{k}}$
averaged with the ``kinetic'' relevant distribution function
\begin{align}    \label{Eq2.12}
 \varrho_{\text{rel}}^{\text{kin-hyd}} \big(x^{N};t\big)&=\exp\left[-\Phi^{\text{kin-hyd}}(t)-
 \int \rd x\gamma(x;t)\hat{n}_{1}(x)-\int \rd{\bf r}
     \beta({\bf r};t)\hat{H}^{\text{int}}({\bf r})\right],               \nonumber \\
 \Phi^{\text{kin-hyd}}(t)&=\ln \int \rd\Gamma_{N}\exp \left[-\int \rd x \gamma (x;t)\hat{n}_{1}(x)
   -\int \rd{\bf r} \beta({\bf r};t)\hat{H}^{\text{int}}({\bf r})\right].
\end{align}
Here, the entropy
\begin{align}    \label{Eq2.13}
 S(t)=-\left\langle \ln \varrho_{\text{rel}} \big(x^{N};t\big)  \right\rangle^{t}_{\text{rel}}
     &= \Phi (t)+\int \rd x \gamma (x;t)\langle \hat{n}_{1}(x)\rangle^{t}+ \int \rd{\bf r} \beta({\bf r};t)\langle \hat{H}^{\text{int}}({\bf r})\rangle^{t}   \nonumber\\
     &\quad+ \int \rd a f(a;t)\ln \frac{f(a;t)}{W(a;t)}\,.
\end{align}
corresponds to the relevant distribution (\ref{Eq2.10}).
In combination with the self-consistent conditions (\ref{Eq2.9}), it can be considered as
entropy of non-equilibrium state. In accordance with (\ref{Eq2.6}), in order to obtain
the explicit form of non-equilibrium distribution function, it is necessary to disclose the action
of Liouville operator on $\varrho_{\text{rel}}(x^{N};t)$ and the action of the Kawasaki-Gunton
projection operator, which in our case has the following structure according to (\ref{Eq2.10}):
\begin{align}    \label{Eq2.14}
 P_{\text{rel}}(t)\varrho'&=\varrho_{\text{rel}} \big(x^{N};t\big)\int \rd\Gamma_{N}\varrho'+
\int \rd x \frac{\partial\varrho_{\text{rel}}\big(x^{N};t\big)}{\partial\langle\hat{n}_{1}(x)\rangle^{t}}
\left[\int \rd\Gamma_{N}\hat{n}_{1}(x)\varrho'-
  \langle \hat{n}_{1}(x) \rangle^{t}\int \rd\Gamma_{N}\varrho'\right]\nonumber\\
&\quad+\int \rd{\bf r} \frac{\partial\varrho_{\text{rel}}\big(x^{N};t\big)}{\partial\langle
\hat{H}^{\text{int}}({\bf r})\rangle^{t}}
\left[\int \rd\Gamma_{N}\hat{H}^{\text{int}}({\bf r})\varrho'-
  \langle \hat{H}^{\text{int}}({\bf r}) \rangle^{t}\int \rd\Gamma_{N}\varrho'\right]   \nonumber\\
&\quad+\int \rd a \frac{\partial \varrho_{\text{rel}} \big(x^{N};t\big)}{\partial \left[\frac{f(a;t)}{W(a;t)}\right]}\frac{1}{W(a;t)}\left[\int \rd\Gamma_{N}\hat{f}(a)\varrho'-
f(a;t)\int \rd\Gamma_{N}\varrho'\right]                                    \nonumber \\
&\quad+\int \rd x \int \rd a \frac{\partial \varrho_{\text{rel}} \big(x^{N};t\big)}{\partial \left[\frac{f(a;t)}{W(a;t)}\right]}\frac{f(a;t)}{W(a;t)}\frac{\partial \ln W(a;t)}{\partial \langle \hat{n}_{1}(x) \rangle^{t}}
\left[\int \rd\Gamma_{N}\hat{n}_{1}(x)\varrho'-
\langle \hat{n}_{1}(x) \rangle^{t}\int \rd\Gamma_{N}\varrho'\right]     \nonumber \\
&\quad+ \int \rd{\bf r} \int \rd a \frac{\partial \varrho_{\text{rel}} \big(x^{N};t\big)}{\partial \left[\frac{f(a;t)}{W(a;t)}\right]}\frac{f(a;t)}{W(a;t)}\frac{\partial \ln W(a;t)}{\partial \langle \hat{H}^{\text{int}}({\bf r})\rangle^{t}}         \left[\int \rd\Gamma_{N}\hat{H}^{\text{int}}({\bf r})\varrho'-
\langle \hat{H}^{\text{int}}({\bf r}) \rangle^{t}\int \rd\Gamma_{N}\varrho'\right].
\end{align}

Next, we consider the action of Liouville operator on a relevant distribution
function (\ref{Eq2.10}):
\begin{align}    \label{Eq2.15}
\ri L_{N}\varrho_{\text{rel}} \big(x^{N};t\big)&=-\int\rd x \gamma (x;t)\dot{\hat{n}}_{1}(x)\varrho_{\text{rel}} \big(x^{N};t\big)
-\int\rd{\bf r} \beta({\bf r};t)\dot{\hat{H}}^{\text{int}}({\bf r})\varrho_{\text{rel}} \big(x^{N};t\big) \nonumber \\
&\quad+ \left[\ri L_{N}\frac{f(a;t)}{W(a;t)}\Bigg{|}_{a=\hat{a}} \right] \varrho_{\text{rel}}^{\text{kin-hyd}} \big(x^{N};t\big),
\end{align}
where $\dot{\hat{n}}_{1}(x)=\ri L_{N}\hat{n}_{1}(x)$,
$\dot{\hat{H}}^{\text{int}}({\bf r})=\ri L_{N}\hat{H}^{\text{int}}({\bf r})$.
Having used thereafter the relation
\begin{eqnarray}
\nonumber
\ri L_{N}\hat{f}(a)=\ri L_{N}\hat{f}(n_{\bf {k}},\bf {J}_{\bf {k}},
    \varepsilon_{\bf {k}})
=\sum_{\bf k} \Bigg[\frac {\partial }{\partial n_{\bf {k}}}\hat{f}(a) \dot{\hat{n}}_{\bf {k}}+
\frac {\partial }{\partial {\bf J}_{\bf k}}\hat{f}(a) \dot{\hat{ {\bf J}}}_{\bf k}+
\frac{\partial}{\partial\varepsilon_{\bf k}}\hat{f}(a) \dot{\hat{\varepsilon}}_{\bf {k}}\Bigg],
\end{eqnarray}
where
$\dot{\hat{n}}_{\bf k}=\ri L_{N}\hat{n}_{\bf k}\,$,
$\dot{\hat{{\bf J}}}_{\bf  k}=\ri L_{N}\hat{{\bf J}}_{\bf k}\,$,
$\dot{\hat{\varepsilon}}_{\bf k}=\ri L_{N}\hat{\varepsilon}_{\bf k}\,$,
the last expression in (\ref{Eq2.15}) can be rewritten in the following form:
\begin{align}    \label{Eq2.16}
\left[\ri L_{N}\frac{f(a;t)}{W(a;t)}\Bigg|_{a=\hat{a}}\right]\varrho_{\text{rel}}^{\text{kin-hyd}} \big(x^{N};t\big) &=
\, \int\rd a \sum_{\bf k}\Bigg\{\dot{\hat{n}}_{\bf k} W(a;t)
\left[\frac {\partial }{\partial n_{\bf k}} \frac {f(a;t)}{W(a;t)}\right]
+\,\dot{\hat{{\bf J}}}_{\bf k}W(a;t)\left[\frac{\partial}{\partial
{\bf J}_{\bf k}}\frac{f(a;t)}{W(a;t)}\right] \nonumber\\
&\quad+\dot{\hat{\varepsilon}}_{\bf k} W(a;t)\left[\frac{\partial}
{\partial\varepsilon_{\bf k}}\frac{f(a;t)}
{W(a;t)}\right]\Bigg\}\varrho_{\text{L}}\big(x^{N};t\big).
\end{align}
Here, we introduced a new relevant distribution function $\varrho_{\text{L}} (x^{N},a;t)$ with the
microscopic distribution of large-scale collective variables
\begin{equation}    \label{Eq2.17}
\varrho_{\text{L}} \big(x^{N},a;t\big)=\varrho_{\text{rel}}^{\text{kin-hyd}} \big(x^{N};t\big)  \frac{\hat{f}(a)}{ W(a;t)}\,.
\end{equation}
This relevant distribution function is connected with $\varrho_{\text{rel}}(x^{N};t)$
by the relation
\begin{equation}    \label{Eq2.18}
\varrho_{\text{rel}}\big(x^{N};t\big)=\int \rd a f(a;t)\varrho_{\text{L}}\big(x^{N},a;t\big)
\end{equation}
and is obviously normalized to unity
\begin{equation}    \label{Eq2.19}
\int\rd\Gamma_{N}\varrho_{\text{L}}\big(x^{N};t\big)=1.
\end{equation}
Using the relation (\ref{Eq2.17}), the average values with a relevant distribution
are conveniently represented in the following form:
\begin{equation}    \label{Eq2.20}
 \langle \ldots \rangle^t_{\text{rel}} = \int\rd a \langle \ldots \rangle^t_{\text{L}} f(a;t),\qquad
 \langle \ldots \rangle^t_{\text{L}} = \int\rd \Gamma_N \ldots\varrho_{\text{L}}(x^N;t).
\end{equation}
Now, in accordance with (\ref{Eq2.16}) and (\ref{Eq2.17}) we can rewrite the action of
the Liouville operator on $\varrho_{\text{rel}}(x^N;t)$ as follows:
\begin{align}    \label{Eq2.21}
\ri L_N \varrho_{\text{rel}}\big(x^N;t\big)&=-\int\rd a \int\rd x
     \gamma(x;t)\dot{\hat{n}}_1(x)\varrho_{\text{L}}\big(x^N,a;t\big)f(a;t)  
-\int\rd a \int\rd{\bf r} \beta({\bf r};t)\dot{\hat{H}}^{\text{int}}({\bf r})
     \varrho_{\text{L}}\big(x^{N},a;t\big)f(a;t)                        \nonumber \\
&\quad+ \int\rd a \sum\limits_{\bf{k}}
  \Bigg[\dot{\hat{n}}_{\bf k}W(a;t)\frac{\partial}{\partial n_{\bf k}}\frac{f(a;t)}{W(a;t)}+
  W(a;t)\dot{\hat{{\bf J}}}_{\bf k}\cdot\frac{\partial}{\partial{\bf J}_{\bf k}}
  \frac{f(a;t)}{W(a;t)}                                           \nonumber  \\
 &\quad
 + \dot{\hat{\varepsilon}}_{\bf k}W(a;t)\frac{\partial}{\partial\varepsilon_{\bf k}}
  \frac{f(a;t)}{W(a;t)}\Bigg]\varrho_{\text{L}}\big(x^{N},a;t\big).
\end{align}
Substituting this expression into (\ref{Eq2.6}), one obtains, for non-equilibrium
distribution function, the following result:
\begin{align}    \label{Eq2.22}
\varrho \big(x^{N};t\big)&=\int\rd a f(a;t)\varrho_{\text{L}}\big(x^{N},a;t\big)       \nonumber \\
&\quad+\int\rd a \int\rd{\bf r} \int_{-\infty}^{t}\rd t'
     \re^{\epsilon (t'-t)}T_{\text{rel}}(t,t')\big[1-P_{\text{rel}}(t')\big]
\dot{\hat{H}}^{\text{int}}({\bf r})\varrho_{\text{L}} \big(x^{N};t\big)f(a;t')\beta ({\bf r};t')\nonumber\\
   &\quad- \int\rd a \int\rd x  \int_{-\infty}^{t}\rd t'
      \re^{\epsilon (t'-t)}T_{\text{rel}}(t,t')\big[1-P_{\text{rel}}(t')\big]
\dot{\hat{n}}_{1}(x)\varrho_{\text{L}} \big(x^{N},a;t'\big)f(a;t')\gamma (x;t')  \nonumber\\
 &\quad-\int\rd a \sum_{\bf k}  \int_{-\infty}^{t}\rd t'
      \re^{\epsilon (t'-t)}T_{\text{rel}}(t,t')\big[1-P_{\text{rel}}(t')\big]
\Bigg[\dot{\hat{n}}_{\bf k}W(a;t')\frac{\partial}{\partial n_{\bf k}}
\frac{f(a;t')}{W(a;t')}
\nonumber  \\
&\quad+ W(a;t')\dot{\hat{{\bf J}}}_{\bf k}\cdot\frac{\partial}{\partial{\bf J}_{\bf k}}
\frac{f(a;t')}{W(a;t')} + \,\,\dot{\hat{\varepsilon}}_{\bf
k}W(a;t')\frac{\partial}{\partial\varepsilon_{\bf k}}
\frac{f(a;t')}{W(a;t')}\Bigg]\varrho_{\text{L}}\big(x^{N},a;t'\big)
\end{align}
and the corresponding generalized transport equations:
\begin{align}    \label{Eq2.23}
&\left(\frac{\partial}{\partial t}+\frac{\bf p}{m}\cdot\frac{\partial}{\partial \bf r}\right)
f_{1}(x;t)-\int\rd x'\frac{\partial }{\partial \bf r}\Phi(|{\bf r} -{\bf r}'|)
\cdot\left(\frac{\partial }{\partial \bf p}-
\frac{\partial }{\partial {\bf p}'}\right)g_{2}(x,x';t)            \nonumber\\
&= - \int\rd{\bf r}'\int\rd a\int_{-\infty}^{t}\rd t'
  \re^{\epsilon(t'-t)}\phi_{nH}(x,{\bf r}',a;t,t')f(a;t')\beta({\bf r}';t') \nonumber\\
&\quad- \int\rd x'\int\rd a\int_{-\infty}^{t}\rd t'
  \re^{\epsilon(t'-t)}\phi_{nn}(x,x',a;t,t')f(a;t')\gamma (x';t') \nonumber\\
&\quad-\sum_{\bf k}\int\rd a\int_{-\infty}^{t}\rd t' \re^{\epsilon (t'-t)}
\Bigg[\phi_{nj}(x,{\bf k},a;t,t')\cdot \frac {\partial  }{\partial {\bf J}_{\bf k}}
+  \phi_{n\varepsilon}(x,{\bf k},a;t,t')\frac{\partial}{\partial\varepsilon_{\bf k}}\Bigg]
\frac{f(a;t')}{W(a;t')}\,, \quad
\end{align}
\begin{align}
\label{Eq2.24}
\frac{\partial}{\partial t}\langle \hat{H}^{\text{int}}({\bf r})\rangle^{t} &=\langle \dot{\hat{H}}^{\text{int}}({\bf r})\rangle^{t}_{\text{rel}}
-\int\rd{\bf r}'\int\rd a\int_{-\infty}^{t}\rd t'
  \re^{\epsilon(t'-t)}\phi_{HH}({\bf r},{\bf r}',a;t,t')f(a;t')\beta ({\bf r}';t') \nonumber\\
&\quad- \int\rd x'\int\rd a\int_{-\infty}^{t}\rd t'
   \re^{\epsilon(t'-t)}\phi_{Hn}({\bf r},x',a;t,t')f(a;t')\gamma (x';t')            \nonumber \\
&\quad-\sum_{\bf k}\int\rd a\int_{-\infty}^{t} \rd t' \re^{\epsilon (t'-t)}
\Bigg[\phi_{Hj}({\bf r},{\bf k},a;t,t')\cdot \frac {\partial}{\partial{\bf J}_{\bf k}}
+ \phi_{H\varepsilon}({\bf r},{\bf k},a;t,t')\frac{\partial}
{\partial\varepsilon_{\bf k}}\Bigg]
\frac{f(a;t')}{W(a;t')}\,, \qquad
\end{align}
\begin{align}
\label{Eq2.25}
\frac{\partial }{\partial t}f(a;t)&=\sum_{\bf k}
\Bigg[\frac {\partial }{\partial n_{{\bf k}}}v_{n}(a;t)
+\frac {\partial  }{\partial {\bf J}_{\bf k}}\cdot {\bf v}_{j}(a;t)
+\frac {\partial  }{\partial \varepsilon_{\bf k}}v_{\varepsilon}(a;t)\Bigg]f(a;t) \nonumber\\
&= \sum_{\bf k}\frac {\partial  }{\partial {\bf J}_{\bf k}}\cdot
  \int\rd{\bf r}'\int\rd a'\int_{-\infty}^{t}\rd t' \re^{\epsilon (t'-t)}
\phi_{jH}({\bf r}',{\bf k},a,a';t,t')f(a;t')\beta ({\bf r}';t')   \nonumber\\
&\quad- \sum_{\bf k}\frac {\partial  }{\partial \varepsilon_{\bf k}}\int\rd{\bf r}'
  \int\rd a'\int_{-\infty}^{t}\rd t' \re^{\epsilon(t'-t)}\phi_{\varepsilon H}({\bf r}',{\bf k},a,a';t,t')f(a;t')\beta ({\bf r}';t')  \nonumber \\
&\quad+\sum_{\bf k}\frac{\partial }{\partial{\bf J}_{\bf k}}\cdot\int\rd x'
\int\rd a'\int_{-\infty}^{t}\rd t' \re^{\epsilon(t'-t)}\phi_{jn}(x',{\bf k},a,a';t,t')
f(a;t')\gamma (x';t') \nonumber\\
&\quad-\sum_{\bf k}\frac {\partial  }{\partial \varepsilon_{\bf k}}
  \int\rd x'\int\rd a'\int_{-\infty}^{t}\rd t' \re^{\epsilon(t'-t)}\phi_{\varepsilon n}(x',{\bf k},a,a';t,t')f(a;t')\gamma (x';t') \nonumber\\
&\quad+ \sum_{\bf k,\bf q}\int\rd a'\int_{-\infty}^{t}\rd t'
   \re^{\epsilon(t'-t)}\frac{\partial}{\partial
{\bf J}_{\bf k}}\cdot \phi_{jj}({\bf k},{\bf q},a,a';t,t')
\cdot\frac {\partial  }{\partial {\bf J}_{\bf q}}\frac {f(a;t')}{W(a;t')} \nonumber \\
&\quad+  \sum_{\bf k,\bf q}\int\rd a'\int_{-\infty}^{t}\rd t' \re^{\epsilon (t'-t)}
\frac {\partial}{\partial {\varepsilon}_{\bf k}}
\phi_{\varepsilon \varepsilon}({\bf k},{\bf q},a,a';t,t')
\frac {\partial}{\partial {\varepsilon}_{\bf q}}\frac {f(a;t')}{W(a;t')} \nonumber\\
&\quad+\sum_{\bf k,\bf q}\int\rd a'\int_{-\infty}^{t}\rd t' \re^{\epsilon (t'-t)}
\Bigg[\frac{\partial}{\partial{\bf J}_{\bf k}}\cdot\phi_{j\varepsilon}({\bf k},{\bf q},a,a';t,t')
\frac {\partial}{\partial{\varepsilon}_{\bf q}}  
+ \frac{\partial}{\partial{\varepsilon}_{\bf k}}
\phi_{\varepsilon j}({\bf k},{\bf q},a,a';t,t')\cdot\frac{\partial}{\partial{\bf J}_{\bf q}}\Bigg]
\frac {f(a;t')}{W(a;t')}\,.
\end{align}
The generalized transport equations (\ref{Eq2.23}), (\ref{Eq2.24}) include
a relevant two-particle distribution function of particles $g_{2}(x,x';t)$ :
\begin{eqnarray}    \label{Eq2.26}
g_{2}(x,x';t)=\int\rd\Gamma_{N-2}\varrho_{\text{rel}} \big(x^{N};t\big)=\langle \hat{n}_{1}(x)\hat{n}_{1}(x')  \rangle_{\text{rel}}^{t}                            
= \langle \hat{G}(x,x')  \rangle_{\text{rel}}^{t} = \int\rd a g_{2}^{\text{L}}(x,x';a;t)f(a;t),
\end{eqnarray}
where
$
\hat{G}(x,x')=\hat{n}_{1}(x)\hat{n}_{1}(x')$,
and $g_{2}^{\text{L}}(x,x';a;t)=\int\rd\Gamma_{N-2}\varrho_{\text{L}} (x^{N};a;t)$
is the two-particle relevant distribution function of large-scale collective variables.
The generalized transport kernels
\begin{equation}    \label{Eq2.27}
\phi_{\alpha \beta}(t,t')=\langle I_{\alpha}(t)T_{\text{rel}}(t,t')I_{\beta}(t') \rangle_{\text{L}}^{t'},
\qquad \alpha, \beta =\{n,H,{\bf j},\varepsilon\},
\end{equation}
that describe non-Markovian kinetic and hydrodynamic processes, are non-equilibrium correlation
functions of generalized fluxes $I_{\alpha}, I_{\beta}$:
\begin{equation}    \label{Eq2.28}
\hat{I}_{n}(x;t)=[1-P(t)]\dot{\hat{n}}_{1}(x), \qquad
\hat{I}_{H}({\bf r};t)=[1-P(t)]\dot{\hat{H}}^{\text{int}}({\bf r}),
\end{equation}
\begin{equation}    \label{Eq2.30}
\hat{I}_{\bf j}({\bf k};t)=[1-P(t)]\dot{\hat{\bf J}}_{\bf k}\,, \qquad
\hat{I}_{\varepsilon}({\bf k};t)=[1-P(t)]\dot{\hat{\varepsilon}}_{\bf k}\,.
\end{equation}
Here, $P(t)$ is the generalized Mori operator related to Kawasaki-Gunton projection operator
$P_{\text{rel}}(t)$ by the following relation
\[
P_{\text{rel}}(t)a(x)\varrho_{\text{rel}} \big(x^{N};t\big)=\varrho_{\text{rel}} \big(x^{N};t\big)P(t)a(x).
\]
It should be emphasized that in (\ref{Eq2.26}) the averages are calculated with
a distribution function $\varrho_{\text{L}}(x^{N},a;t)$ (\ref{Eq2.20}), so that the transport kernels
are functions of collective variables $a_{\bf k}$.
In equation (\ref{Eq2.25}), the functions (called hydrodynamic velocities) $v_{n,{\bf k}}(a;t)$, ${\bf v}_{j,{\bf k}}(a;t)$, $v_{\varepsilon,{\bf k}}(a;t)$ represent the fluxes
in the space of collective variables and are defined as:
\begin{eqnarray}        \label{Eq2.32}
\nonumber
 v_{n}(a;t)=\int\rd\Gamma_{N}\dot{\hat{n}}_{\bf k}\varrho_{\text{L}} \big(x^{N},a;t\big)= \langle
\dot{\hat{n}}_{\bf k} \rangle_{\text{L}}^{t},                \\
 v_{j}(a;t)=\int\rd\Gamma_{N}\dot{\hat{{\bf J}}}_{\bf k}\varrho_{\text{L}} \big(x^{N},a;t\big)= \langle
 \dot{\hat{{\bf J}}}_{\bf k} \rangle_{\text{L}}^{t},        \nonumber \\
 v_{\varepsilon}(a;t)=\int\rd\Gamma_{N}\dot{\hat{\varepsilon}}_{\bf k}\varrho_{\text{L}} \big(x^{N},a;t\big)=
 \langle \dot{\hat{\varepsilon}}_{\bf k} \rangle_{\text{L}}^{t}.
\end{eqnarray}
The presented system of transport equations provides a consistent description of kinetic and
hydrodynamic processes of classical fluids which take into account long-living fluctuations.

The system of transport equations (\ref{Eq2.23})--(\ref{Eq2.25}) is not closed
due to Lagrange parameters $\gamma(x;t)$, $\beta ({\bf r}';t')$,
which are determined from the corresponding self-consistent conditions.

To describe the kinetics of heterophase fluctuations, as discussed in the Introduction,
in addition to the equation for non-equilibrium one-particle distribution function
it is important to consider the equations for higher non-equilibrium distribution functions
of groups of interacting particles. The kinetics of these phases is described by
kinetic equations for higher non-equilibrium distribution functions of finite number
of particles in heterophase formations.
Therefore, to describe these heterophase  kinetic processes  we must supplement
this system of transport equations with the kinetic equation
$f_{2}(x,x';t)$, and hence for $f_{s}(x_1 \dots x_s;t)$,  $s<N$:
\begin{align}    \label{Eq2.33}
&\frac{\partial}{\partial t}
f_{2}(x,x';t)+\ri L_{2}f_{2}(x,x';t)
-\int\rd x''[\ri L(13)+\ri L(23)]f_{3}(x,x',x'';t)      \nonumber\\
&= \ri L_{2}\Delta f_{2}(x,x';t)-\int dx''[\ri L(13)+\ri L(23)]
    \Delta f_{3}(x,x',x'';t)                         \nonumber \\
&\quad+ \int\rd x''\int\rd a\int_{-\infty}^{t} \rd t'
\re^{\epsilon(t'-t)}\phi_{Gn}(x,x',x'',a;t,t')f(a;t')\gamma (x'';t')  \nonumber\\
&\quad-\int\rd{\bf r}''\int\rd a\int_{-\infty}^{t}\rd t'
 \re^{\epsilon(t'-t)}\phi_{GH}(x,x',{\bf r}'',a;t,t')f(a;t')\beta ({\bf r}'';t')  \nonumber \\
&\quad-\sum_{\bf k}\int\rd a\int_{-\infty}^{t}\rd t'\re^{\epsilon (t'-t)}
\Bigg[\phi_{Gj}(x,x',{\bf k},a;t,t')\cdot \frac {\partial}{\partial{\bf J}_{\bf k}}
+ \, \phi_{G\varepsilon}(x,x',{\bf k},a;t,t')\frac{\partial}{\partial\varepsilon_{\bf k}}\Bigg]
\frac{f(a;t')}{W(a;t')}\,,
\end{align}
\begin{align}    \label{Eq2.34}
&\frac{\partial}{\partial t}
f_{s}(x^{s};t)+\ri L_{s}f_{s}(x^{s};t)-\sum_{j}\frac {1}{s!}\int\rd x_{s+1}
  \ri L(j,s+1)f_{s+1}(x^{s},x_{s+1};t)       \nonumber \\
&= \ri L_{s}\Delta f_{s}(x^{s};t)-\sum_{j}\frac {1}{s!}\int dx_{s+1}
     \ri L(j,s+1)\Delta f_{s+1}(x^{s},x_{s+1};t)                       \nonumber \\
&\quad+\int\rd x''\int\rd a\int_{-\infty}^{t}\rd t'
  \re^{\epsilon(t'-t)}\phi_{G_{s}n}(x^{s},x'',a;t,t')f(a;t')\gamma (x'';t') \nonumber  \\
&\quad- \int\rd{\bf r}''\int\rd a\int_{-\infty}^{t}\rd t'
  \re^{\epsilon(t'-t)}\phi_{G_{s}H}(x^{s},{\bf r}'',a;t,t')f(a;t')\beta ({\bf r}'';t') \nonumber\\
&\quad- \sum_{\bf k}\int\rd a\int_{-\infty}^{t}\rd t' \re^{\epsilon (t'-t)}
\Bigg[\phi_{G_{s}j}(x^{s},{\bf k},a;t,t')\cdot \frac {\partial  }{\partial {\bf J}_{\bf k}}
+ \phi_{G_{s}\varepsilon}(x^{s},{\bf k},a;t,t')\frac{\partial}{\partial\varepsilon_{\bf k}}\Bigg]
\frac{f(a;t')}{W(a;t')}\,,
\end{align}
where \[\Delta f_{2}(x,x';t)=f_{2}(x,x';t)-g_{2}(x,x';t),\qquad
   \Delta f_{s}(x^{s};t)=f_{s}(x^{s};t)-g_{s}(x^{s};t).  \]
In equation (\ref{Eq2.33}), the two-particle Liouville operator
\[\ri L_{2}=\ri L_{0}(x)+iL_{0}(x')+\ri L(x,x')\]
was introduced. It contains a one-particle operator
\[\ri L_{0}(x)=\frac{\bf p}{m}\cdot\frac{\partial}{\partial \bf r}\,, \qquad x=\{{\bf r,p}\},\]
as well as a potential part
\[
\ri L(x,x')=\frac{\partial }{\partial \bf r}\Phi(|{\bf r} -{\bf r}'|)
\cdot\Bigg(\frac{\partial }{\partial \bf p}-
\frac{\partial }{\partial {\bf p}'}\Bigg).
\]
Accordingly, in equation (\ref{Eq2.34}), $\ri L_{s}$ is the $s$-particle Liouville operator,
\[
g_{s}(x_1\dots x_{s};t)=\langle \hat{G}_{s}(x_1\dots x_{s}) \rangle_{\text{rel}}^{t}=
\int\rd a g_{s}^{\text{L}}(x_1\dots x_{s};a;t)f(a;t),
\]
where
\[
g_{s}^{\text{L}}(x_1 \dots x_{s};a;t)=\int\rd\Gamma_{N}\hat{G}_{s}(x_1\dots x_{s})\varrho_{\text{L}}\big(x^{N};a;t\big)
\]
is the $s$-particle relevant distribution function of large-scale variables
and $\hat{G}_{s}(x^{s})=\hat{n}_{1}(x_{1})\ldots\hat{n}_{1}(x_{s})$.

Thus, we obtained a system of equations for non-equilibrium
one-, two-, $s$-particle distribution functions
which take into account nonlinear hydrodynamic fluctuations.

Now, we discuss the equation (\ref{Eq2.25}) that is of Fokker-Planck type for a non-equilibrium
distribution function of collective variables which take into account the kinetic processes.
The transport kernel in this equation $\phi_{nn}(x,x';t,t')$ describes a dissipation
of kinetic processes, while the kernels
$\phi_{nj}(x,{\bf k},a;t,t')$, $\phi_{n\varepsilon}(x,{\bf k},a;t,t')$,
$\phi_{jn}(x,{\bf k},a;t,t')$, $\phi_{\varepsilon n}(x,{\bf k},a;t,t')$ describe
a dissipation of correlations between kinetic and hydrodynamic processes.
To uncover a more detailed structure  of transport
kernels $\phi _{nn}(x;x',a;t,t')$, $\phi _{Gn} (x;x',x'',a;t,t')$
we consider the action of Liouville operator on $\hat{n}_{1} (x)$ and  $\hat{G} (x,x')$:
\begin{eqnarray}            \label{Eq2.35}
\ri L_{N} \hat{n}_{1} (x)=-\frac{\partial }{\partial {\bf r}} \cdot
\frac{1}{m} \hat{{\bf j}}({\bf r},{\bf p})+\frac{\partial }{\partial {\bf p}} \cdot
\hat{{\bf F}}({\bf r},{\bf p}),
\end{eqnarray}
\begin{eqnarray}            \label{Eq2.36}
\ri L_{N} \hat{G}(x,x')=-\frac{\partial }{\partial {\bf r}} \cdot
\frac{1}{m} \hat{{\bf j}}({\bf r},{\bf p})\hat{n}_{1} (x')-\hat{n}_{1} (x)
\frac{\partial }{\partial {\bf r}'} \cdot \frac{1}{m} \hat{{\bf j}}({\bf r}',{\bf p}')
+\frac{\partial }{\partial {\bf p}} \cdot \hat{{\bf F}}({\bf r},{\bf p})\hat{n}_{1} (x')+\hat{n}_{1} (x)\frac{\partial }{\partial {\bf p}'} \cdot \hat{{\bf F}}({\bf r}',{\bf p}'),
\end{eqnarray}
where
\begin{equation}           \label{Eq2.37}
\hat{{\bf j}}({\bf r},{\bf p})=\sum _{j=1}^{N} {\bf p}_{j} \delta ({\bf r}-{\bf r}_{j} )\delta ({\bf p}-{\bf p}_{j} )
\end{equation}
is the microscopic density of the momentum vector in coordinate-momentum space,
\begin{equation}            \label{Eq2.38}
\hat{{\bf F}}({\bf r},{\bf p})=\sum _{l\ne j} \frac{\partial }{\partial {\bf r}_{j} } \Phi (|{\bf r}_{j} -{\bf r}_{l} |)\delta ({\bf r}-{\bf r}_{j} )\delta ({\bf p}-{\bf p}_{j} )
\end{equation}
is the microscopic density of force vector in coordinate-momentum space.
Taking into account the equations (\ref{Eq2.35})--(\ref{Eq2.38})
for the kinetic transport kernels, we obtain:
\begin{align}             \label{Eq2.39}
\phi _{nn} (x;x',a;t,t')&=
-\Bigg[\frac{\partial}{\partial{\bf r}}\cdot D_{jj} (x,x',a;t,t')\cdot\
\frac{\partial }{\partial {\bf r}'}
- \frac{\partial}{\partial{\bf p}}\cdot D_{Fj}(x,x',a;t,t')\cdot
\frac{\partial}{\partial{\bf r}'}  \nonumber         \\
&\quad-\frac{\partial }{\partial {\bf r}} \cdot D_{jF} (x,x',a;t,t')\cdot
\frac{\partial }{\partial {\bf p}'} +\frac{\partial }{\partial {\bf p}} \cdot D_{FF}
(x,x',a;t,t')\cdot \frac{\partial }{\partial {\bf p}'} \Bigg],
\end{align}
where
\begin{equation}         \nonumber
D_{jj}(x,x',a;t,t')=\int\rd\Gamma _{N} \hat{{\bf j}}(x)T(t,t')[1-P(t')]
    \hat{{\bf j}}(x')\rho_{\text{L}}\big(x^{N};t'\big),
\end{equation}
\begin{equation}         \nonumber
D_{FF}(x,x',a;t,t')=\int\rd\Gamma_{N}\hat{{\bf F}}(x)T(t,t')[1-P(t')]
    \hat{{\bf F}}(x')\rho_{\text{L}}\big(x^{N};t'\big)
\end{equation}
are the generalized diffusion and the particle friction coefficients
in the coordinate-momentum space. Moreover,
\begin{equation}         \nonumber
\int\rd{\bf p}\int\rd{\bf p}'  D_{jj}(x,x';t,t')=D_{jj} ({\bf r},{\bf r}';t,t'), \qquad
\int\rd{\bf p}\int\rd{\bf p}'  D_{FF}(x,x';t,t')=D_{FF}({\bf r},{\bf r}';t,t')
\end{equation}
are the generalized coefficients of diffusion and friction. Similarly, we obtain
the expression for the transport kernel $\phi_{Gn}(x;x',x'';t,t')$:
\begin{align}          \label{Eq2.40}
\phi _{Gn} (x;x',x'',a;t,t')&=
-\Bigg[\frac{\partial }{\partial {\bf r}} \cdot D_{jjn} (x,x',x'',a;t,t')
\cdot \frac{\partial}{\partial {\bf r}'}
+\frac{\partial }{\partial {\bf r}} \cdot D_{jnj} (x,x',x'',a;t,t')\cdot
\frac{\partial }{\partial {\bf r}''}\nonumber     \\
&\quad-\frac{\partial }{\partial {\bf p}} \cdot D_{Fjn} (x,x',x'',a;t,t')\cdot
\frac{\partial }{\partial {\bf r}'}
-\frac{\partial }{\partial {\bf p}} \cdot D_{Fnj} (x,x',x'',a;t,t')\cdot
\frac{\partial }{\partial {\bf r}''}\nonumber \\
&\quad-\frac{\partial }{\partial {\bf r}} \cdot D_{jFn} (x,x',x'',a;t,t')\cdot
\frac{\partial }{\partial {\bf p}'}
-\frac{\partial }{\partial {\bf r}} \cdot D_{jnF} (x,x',x'',a;t,t')\cdot
\frac{\partial }{\partial {\bf p}''}\nonumber     \\
 &\quad+\frac{\partial }{\partial {\bf p}} \cdot D_{FFn} (x,x',x'',a;t,t')\cdot
 \frac{\partial }{\partial {\bf p}'}
+\frac{\partial }{\partial {\bf p}} \cdot D_{FnF} (x,x',x'',a;t,t')\cdot
\frac{\partial }{\partial {\bf p}''}\Bigg].
\end{align}
It is remarkable that the expression
\[\int\rd x'\int\rd a\int_{-\infty}^{t}\rd t'
\re^{\epsilon(t'-t)}\phi_{nn}(x,x',a;t,t')f(a;t')\gamma (x';t')\]
in equation (\ref{Eq2.23}) with (\ref{Eq2.39}) is the generalized collision integral of
Fokker-Planck type in the coordinate-momentum space.
That is, taking into account (\ref{Eq2.26}) and (\ref{Eq2.39}), the kinetic equation
(\ref{Eq2.23}) can be written as follows:
\begin{align}
&\Bigg(\frac{\partial}{\partial t}+\frac{\bf p}{m}
\frac{\partial}{\partial \bf r}\Bigg)
f_{1}(x;t)-\int\rd x'\int\rd a\frac{\partial }{\partial \bf r}\Phi(|{\bf r}-{\bf r}'|)
\Bigg(\frac{\partial }{\partial \bf p}-
\frac{\partial }{\partial {\bf p}'}\Bigg)g_{2}^{l}(x,x',a;t)f(a;t)  \nonumber\\
&=-\int\rd{\bf r}'\int\rd a\int_{-\infty}^{t}\rd t' \re^{\epsilon(t'-t)}
\phi_{nH}(x,{\bf r}',a;t,t')f(a;t')\beta ({\bf r}';t')        \nonumber  \\
&\quad-\int\rd x'\int\rd a\int_{-\infty}^{t}\rd t'
   \re^{\epsilon(t'-t)}\frac{\partial }{\partial {\bf r}} \cdot
   D_{jj} (x,x',a;t,t')\cdot \frac{\partial }{\partial {\bf r}'}\gamma (x';t')f(a;t')      \nonumber \\
 &\quad+\int\rd x'\int\rd a\int_{-\infty}^{t}\rd t'\re^{\epsilon(t'-t)}
\Bigg[ \frac{\partial }{\partial {\bf p}} \cdot D_{Fj} (x,x',a;t,t')\cdot
\frac{\partial }{\partial {\bf r}'}               \nonumber   \\
  &\quad+\frac{\partial }{\partial {\bf r}} \cdot D_{jF} (x,x',a;t,t')\cdot
\frac{\partial }{\partial {\bf p}'} -\frac{\partial }{\partial {\bf p}} \cdot
D_{FF}(x,x',a;t,t')\cdot \frac{\partial}{\partial {\bf p}'} \Bigg]
\gamma (x';t')f(a;t')   \nonumber      \\
&\quad-\sum_{\bf k}\int\rd a\int_{-\infty}^{t}\rd t'\re^{\epsilon (t'-t)}
\Bigg[\phi_{nj}(x,{\bf k},a;t,t')\cdot \frac {\partial  }{\partial {\bf J}_{\bf k}}
+\phi_{n\varepsilon}(x,{\bf k},a;t,t')\frac{\partial}{\partial\varepsilon_{\bf k}}\Bigg]
\frac{f(a;t')}{W(a;t')}\,.
\label{Eq2.41}
\end{align}
In the equation (\ref{Eq2.25}), the quantities $\phi_{jj}({\bf k},{\bf q},a,a';t,t')$,
$\phi_{j \varepsilon}({\bf k},{\bf q},a,a';t,t')$, $\phi_{\varepsilon j}({\bf k},{\bf
q},a,a';t,t')$, \linebreak $\phi_{\varepsilon \varepsilon}({\bf k},{\bf q},a,a';t,t')$ correspond to
the dissipative processes connected with the correlations between viscous and heat
hydrodynamic processes.
Note that the equations (\ref{Eq2.33}), (\ref{Eq2.34}) have the structure similar to the equation
(\ref{Eq2.41}) but with more complex many-particle transport kernels.
Thus, we obtained a system of equations for non-equilibrium one-, two-,
$s$-particle distribution functions  on collision integral of Fokker-Planck type
which take into account nonlinear hydrodynamic fluctuations.

The set of equations (\ref{Eq2.23}), (\ref{Eq2.25}), (\ref{Eq2.33}), (\ref{Eq2.34})
allow for two limiting cases.
The first, if the description of non-equilibrium processes does not take into account
nonlinear hydrodynamic fluctuations, we will obtain the system of equations
for a consistent description of kinetics and hydrodynamics obtained in papers
\cite{ZubMorOmTok1993,TokOmelKobr1998,MarkOmelTok2014}.
These systems can be written as follows:
\begin{align}    \label{Eq2.42}
&\Bigg(\frac{\partial}{\partial t}+\frac{\bf p}{m}\cdot\frac{\partial}{\partial \bf r}\Bigg)
f_{1}(x;t)-\int\rd x'\frac{\partial }{\partial \bf r}\Phi(|{\bf r} -{\bf r}'|)
\cdot\Bigg(\frac{\partial }{\partial \bf p}-
\frac{\partial }{\partial {\bf p}'}\Bigg)g_{2}(x,x';t) \qquad \nonumber\\
&= - \int\rd{\bf r}' \int_{-\infty}^{t}\rd t'\re^{\epsilon(t'-t)}
\phi_{nH}(x,{\bf r}';t,t')\beta ({\bf r}';t')
- \int\rd x' \int_{-\infty}^{t}\rd t' \re^{\epsilon(t'-t)}
\phi_{nn}(x,x';t,t')\gamma (x';t') ,
\end{align}
\begin{align}
\label{Eq2.43}
\frac{\partial}{\partial t}\langle \hat{H}^{\text{int}}({\bf r})\rangle^{t}
&=\langle \dot{\hat{H}}^{\text{int}}({\bf r})\rangle^{t}_{\text{rel}}
-\int\rd{\bf r}' \int_{-\infty}^{t}\rd t' \re^{\epsilon(t'-t)}
\phi_{HH}({\bf r},{\bf r}';t,t')\beta ({\bf r}';t') \nonumber\\
&\quad- \int\rd x' \int_{-\infty}^{t}\rd t' \re^{\epsilon(t'-t)}
\phi_{Hn}({\bf r},x';t,t') \gamma (x';t'),\qquad\qquad
\end{align}
where the transport kernels
$\phi_{nn}(x,x';t,t')$, $\phi_{nH}(x,{\bf r}';t,t')$, $\phi_{Hn}({\bf r},x';t,t')$,
$\phi_{HH}({\bf r},{\bf r}';t,t')$ are given by expression (\ref{Eq2.27}).
An averaging in these kernels is performed  using
the relevant statistical operator $\varrho^{\text{kin-hyd}}_{\text{rel}}(x^{N};t)$.
If in equations (\ref{Eq2.42}), (\ref{Eq2.43}) we do not consider the contribution
from the potential energy of interaction, which is true for a dilute gas, then we shall obtain
a generalized kinetic equation for a non-equilibrium one-particle distribution function
\cite{Cohen1962}:
\begin{align}       \label{Eq2.44}
 &\Bigg(\frac{\partial}{\partial t}+\frac{\bf p}{m}\cdot\frac{\partial}{\partial\bf r}\Bigg)
  f_{1}(x;t)-\int\rd x'\frac{\partial }{\partial \bf r}\Phi(|{\bf r} -{\bf r}'|)\Bigg(\frac{\partial }{\partial \bf p}-
 \frac{\partial }{\partial {\bf p}'}\Bigg)g_{2}(x,x';t)   \nonumber\\
&
= \int\rd x'\int\limits_{-\infty}^{t}\rd t'
 \re^{\epsilon (t'-t)}\phi_{nn}(x,x';t,t')\gamma (x';t').
\end{align}

The second, if we do not take into account the kinetic processes, then we shall obtain
generalized (non-Markov) Fokker–Planck equation for non-equilibrium distribution function
of collective variables, which can be obtained using the method of Zwanzig
projection operators or using the method of Zubarev non-equilibrium statistical
operator \cite{ZubarMoroz1983}:
\begin{align}    \label{Eq2.45}
\frac{\partial }{\partial t}f(a;t)&=\sum_{\bf k}\Bigg[\frac {\partial }
{\partial n_{{\bf k}}}v_{n}(a;t)+ \frac {\partial  }{\partial {\bf J}_{\bf k}}\cdot
{\bf v}_{j}(a;t)+
\frac {\partial }{\partial \varepsilon_{\bf k}}v_{\varepsilon}(a;t)\Bigg]f(a;t)   \nonumber \\
&= \sum_{\bf k,\bf q}\int\rd a'\int\limits_{-\infty}^{t}\rd t' \re^{\epsilon (t'-t)}
\frac {\partial  }{\partial {\bf J}_{\bf k}}\cdot \phi_{jj}({\bf k},{\bf q},a,a';t,t')
\cdot\frac {\partial  }{\partial {\bf J}_{\bf q}}\frac {f(a;t')}{W(a;t')}     \nonumber\\
&\quad+  \sum_{\bf k,\bf q}\int\rd a'\int\limits_{-\infty}^{t}\rd t' \re^{\epsilon (t'-t)}
\frac {\partial  }{\partial {\varepsilon}_{\bf k}}
\phi_{\varepsilon \varepsilon}({\bf k},{\bf q},a,a';t,t')
\frac {\partial  }{\partial {\varepsilon}_{\bf q}}\frac {f(a;t')}{W(a;t')}
+ \sum_{\bf k,\bf q}\int\rd a'\int\limits_{-\infty}^{t}\rd t' \re^{\epsilon (t'-t)}\nonumber\\
&\quad\times\Bigg[\frac {\partial  }{\partial {\bf J}_{\bf k}}
 \cdot \phi_{j \varepsilon}({\bf k},{\bf q},a,a';t,t')
 \frac {\partial}{\partial {\varepsilon}_{\bf q}}
+  \frac {\partial}{\partial {\varepsilon}_{\bf k}}
\phi_{\varepsilon j}({\bf k},{\bf q},a,a';t,t')
\cdot\frac{\partial}{\partial{\bf J}_{\bf q}}\Bigg]
\frac {f(a;t')}{W(a;t')}\,.
\end{align}

One of the main problems for the analysis of transport equations (\ref{Eq2.23})--(\ref{Eq2.25})
and transport kernels are the calculations of the structure functions $W(a;t)$
of collective variables and of hydrodynamic velocities $v_{n}(a;t)$, $v_{j}(a;t)$,
$v_{\varepsilon}(a;t)$.
We  consider one of these ways  in the next section.

\section{Calculation of structure function $W(a;t)$ and hydrodyna\-mical
         velocities $v_{\alpha}(a;t)$ by the collective variable method}
\label{sec3}

In the Kawasaki theory \cite{Kawasaki1976} of non-linear fluctuations,
the structure function is approximated by a Gaussian dependence on collective variables.
In this case, as it can be seen, the hydrodynamic velocities
$v_{l,{\bf k}}(a;t)$, $l=n,j,\varepsilon$ are the linear function of $a$.
Another approach for the calculation of hydrodynamical velocities $v_{l,{\bf k}}(a;t)$
was proposed on the basis of local thermodynamics \cite{ZubarMoroz1983}.
The resulting expressions  are obviously valid at low frequencies and
for small values of the wave vector, when the conditions of the local thermodynamics
are valid.
Structure function $W(a;t)$ and hydrodynamical velocities $v_{l,{\bf k}}(a;t)$
in a case of study of hydrodynamic fluctuations were calculated in
\cite{Zubarev1982,IdzIgnTok1996} using the method of collective variables \cite{YukhnHolov1980}.
The basic idea of this approach is that the structure function $W(a;t)$ and hydrodynamic
velocities $v_{l,{\bf k}}(a;t)$ can be calculated in approximations higher than Gaussian.
Next, we use the method of collective variables
\cite{Zubarev1982,YukhnHolov1980,IdzIgnTok1996,HlushTokar2016}
to calculate the structure function $W(a;t)$ and hydrodynamic velocities $v_{l,\bf k}(a;t)$.
Further, the collective variable method is used to calculate
the structure function $W(a;t)$ and hydrodynamic velocities $v_{l,{\bf k}}(a;t)$.
We considered earlier the case when the interactions between particles on
small distances were described by short-range potential $\Phi^{\text{sh}}(|{\bf r}_{l j}|)$,
in particular, by the potential of hard spheres.
At long distances, the interactions between particles are described
by the long-range potential $\Phi^{\text{long}}(|{\bf r}_{l j}|)$.
Accordingly, we define the short- and long-acting
parts of the interaction of the Liouville operator:
\[
  \ri L_{N} = \ri L_{N}^{0} +\hat{T}_{N} + \ri L_{N}^{\text{long}},
\]
where $\ri L_{N}^{0}$ is the  Liouville operator of $N$ non-interacting particles,
$\hat{T}_{N}$ is the scattering operator of hard sphere system and
$\ri L_{N}^{\text{long}}$ is the potential part of Liouville operator with long-range interaction
between particles.

First, we calculate the structure function $W(a;t)$ for collective variables.
To do this, we use the integral representation for $\delta$-functions:
\begin{equation}    \label{Eq3.1}
\hat{f}(a)=\int \rd\omega \exp \Bigg[ -\ri\pi \sum_{l,\bf k}
\omega_{l,\bf k}(\hat{a}_{l,\bf k}-a_{l,\bf k})\Bigg],\qquad
l=n,{\bf j},\varepsilon.
\end{equation}
Next, using a cumulant expansion \cite{IdzIgnTok1996,HlushTokar2016} for $W(a;t)$, one obtains:
\begin{align}    \label{Eq3.2}
W(a;t)&=\int\rd\Gamma_{N}\varrho_{\text{rel}}^{\text{kin-hyd}} \big(x^{N};t\big)\hat{f}(a)
=\int\rd\omega \exp \Bigg[-\ri\pi \sum_{l,\bf k}\omega_{l,\bf k}\bar{a}_{l,\bf k}-
\frac{1}{2V^2}\sum_{\bf q}\sum_{\bf k}\beta_{-{\bf q}}(t)\nu({\bf k})(n_{{\bf q}+
{\bf k}}n_{-{\bf k}}-n_{{\bf q}})
 \nonumber \\
&\quad-\frac{\pi^{2}}{2}\sum_{l_{1},l_{2}}\sum_{{\bf k}_{1},{\bf k}_{2}}
\mathfrak{M}_{2}^{l_{1},l_{2}}({\bf k}_{1},{\bf k}_{2};t)\omega_{l_{1},{\bf k}_{1}}
 \omega_{l_{2},{\bf k}_{2}}\Bigg]\exp\Bigg[ \sum_{n\geqslant 3}D_{n}(\omega ;t)\Bigg],
\end{align}
where
\begin{eqnarray}     \nonumber
\bar{a}_{l,\bf k}=a_{l,\bf k}-\langle\hat{a}_{l,\bf k}\rangle_{\text{kin-sh}}^{t}\,, \qquad
 \rd\omega=\prod_{l,\bf k}\rd\omega_{l,\bf k}^{r}\rd\omega_{l,\bf k}^{s}\,,   \qquad
 \omega_{l,\bf k}=\omega_{l,\bf k}^{r}-\ri\omega_{l,\bf k}^{s}\,, \qquad
 \omega_{l,-\bf k}=\omega_{l,\bf k}^{*}\,,
\end{eqnarray}
\begin{equation}    \label{Eq3.3}
 D_{n}(\omega ;t)=\frac{(-\ri\pi)^{n}}{n!}\sum_{l_{1},\ldots,l_{n}}
 \sum_{{\bf k}_{1},\ldots,{\bf k}_{n}}
 \mathfrak{M}_{n}^{l_{1},\ldots,l_{n}}({\bf k}_{1},\ldots,{\bf k}_{n};t)
 \omega_{l_{1},{\bf k}_{1}}\ldots  \omega_{l_{n},{\bf k}_{n}},
\end{equation}
\begin{equation}    \label{Eq3.4}
\mathfrak{M}_{n}^{l_{1},\ldots,l_{n}}({\bf k}_{1},\ldots,{\bf k}_{n};t)=
\langle \hat{a}_{l_{1},{\bf k}_{1}}\ldots\hat{a}_{l_{n},{\bf k}_{n}}\rangle_{\text{kin-sh}}^{t,\text c}
\end{equation}
are the non-equilibrium cumulant averages in approximations of the $n$-order,
which are calculated using distribution $\varrho_{\text{rel}}^{\text{kin-sh}}(x^{N};t)$
for hard sphere model:
\begin{equation}    \label{Eq3.5}
 \varrho_{\text{rel}}^{\text{kin-sh}} \big(x^{N};t\big)=\exp\Bigg[-\Phi^{\text{kin-sh}}(t)-\int \rd{\bf r}
  \beta({\bf r};t)\hat{H}^{\text{sh}}({\bf r})-\int\rd x \gamma (x;t)\hat{n}_{1}(x)\Bigg].
\end{equation}
In (\ref{Eq3.4}) superscript ``c'' means the cumulant averages.
It should be noted that in (\ref{Eq3.2}) we shared the contributions from
short- and long-range interactions.
The short-range interactions are take into account in a relevant distribution
(\ref{Eq3.5}) (which can be considered as a base distribution)
and long-range interactions are presented through collective variables
\[\int \rd{\bf r}
  \beta({\bf r};t)\hat{H}^{\text{long}}({\bf r})=\frac{1}{2V^2}\sum_{\bf q}\sum_{\bf k}
  \beta_{-{\bf q}}(t)\nu({\bf k})\left(n_{{\bf q}+{\bf k}}n_{-{\bf k}}
  - n_{{\bf q}}\right).\]

We present the structure function $W(a;t)$ for further calculations in the following form:
\begin{align}    \label{Eq3.6}
 W(a;t)&=\int\rd\omega \exp\Bigg[ -\ri\pi \sum_{l,\bf k}\omega_{l,\bf k}\bar{a}_{l,\bf k}
- \frac{1}{2V^2}\sum_{\bf q}\sum_{\bf k}\beta_{-{\bf q}}(t)\nu({\bf k})\left(n_{{\bf q}+{\bf k}}
n_{-{\bf k}} - n_{{\bf q}}\right)           \nonumber   \\
&\quad -\frac{\pi^{2}}{2}\sum_{l_{1},l_{2}}\sum_{{\bf k}_{1},{\bf k}_{2}}
 \mathfrak{M}_{2}^{l_{1},l_{2}}({\bf k}_{1},{\bf k}_{2};t)\omega_{l_{1},{\bf k}_{1}}
 \omega_{l_{2},{\bf k}_{2}}\Bigg]
 \Bigg(1+B+\frac{1}{2!}B^{2}+\frac{1}{3!}B^{3}+\ldots+\frac{1}{n!}B^{n}+\ldots\Bigg),
\end{align}
where $B=\sum_{n\geqslant 3}D_{n}(\omega ;t)$. If in series of exponent (\ref{Eq3.6}), namely,
$\exp\big[\sum_{n\geqslant 3}D_{n}(\omega;t)\big]$, one retains only the first term equal to unity,
one will obtain the Gaussian approximation for $W(a;t)$:
\begin{align}    \label{Eq3.7}
W^{\text{G}}(a;t)&=\int\rd\omega \exp \Bigg[ \ri\pi \sum_{l,\bf k}\omega_{l,\bf k}\bar{a}_{l,\bf k}
- \,\frac{1}{2V^2}\sum_{\bf q}\sum_{\bf k}\beta_{-{\bf q}}(t)\nu({\bf k})\left(n_{{\bf q}+{\bf k}}n_{-{\bf k}} -
   n_{{\bf q}}\right) \nonumber\\
&\quad-  \frac{\pi^{2}}{2}\sum_{l_{1},l_{2}}\sum_{{\bf k}_{1},{\bf k}_{2}}
  \mathfrak{M}_{2}^{l_{1},l_{2}}({\bf k}_{1},{\bf k}_{2};t)
  \omega_{l_{1},{\bf k}_{1}} \omega_{l_{2},\,{\bf k}_{2}}\Bigg],
\end{align}
 where $\mathfrak{M}_{2}^{l_{1},l_{2}}({\bf k}_{1},{\bf k}_{2};t)$ are the matrix elements of
non-equilibrium correlation functions:
\begin{equation}    \label{Eq3.8}
\mathfrak{M}_{2}({\bf k}_{1},{\bf k}_{2};t)=
\begin{vmatrix}
 \langle \hat{n}\hat{n}\rangle_{\text{kin-sh}}^{\text{c}}&
 \langle \hat{n}\hat{{\bf J}}\rangle_{\text{kin-sh}}^{\text{c}} &
 \langle \hat{n}\hat{\varepsilon}\rangle_{\text{kin-sh}}^{\text{c}}   \vspace{1mm}\\ \vspace{1mm}
  \langle \hat{{\bf J}}\hat{n}\rangle_{\text{kin-sh}}^{\text{c}}        &
  \langle \hat{{\bf J}}\hat{{\bf J}}\rangle_{\text{kin-sh}}^{\text{c}}  &
  \langle \hat{{\bf J}}\hat{\varepsilon}\rangle_{\text{kin-sh}}^{\text{c}}  \\
  \langle \hat{\varepsilon}\hat{n}\rangle_{\text{kin-sh}}^{\text{c}}  &
  \langle \hat{\varepsilon}\hat{{\bf J}}\rangle_{\text{kin-sh}}^{\text{c}}  &
  \langle \hat{\varepsilon}\hat{\varepsilon}\rangle_{\text{kin-sh}}^{\text{c}}
\end{vmatrix}_{{\bf k_1,k_2}},
\end{equation}
and, for example, the non-equilibrium cumulant average
\begin{equation}    \label{Eq3.9}
\langle \hat{n}_{\bf k} \hat{n}_{-\bf k}\rangle_{\text{kin-sh}}^{t,\,\text{c}}=
\langle \hat{n}_{\bf k} \hat{n}_{-\bf k}\rangle_{\text{kin-sh}}^{t}-
\langle \hat{n}_{\bf k} \rangle_{\text{kin-sh}}^{t}\langle  \hat{n}_{-\bf k}\rangle_{\text{kin-sh}}^{t}\,.
\end{equation}
For integrating over $\rd\omega$ in (\ref{Eq3.7}), we should transform the quadratic form
in an exponential expression into a diagonal form with respect to $\omega_{l,{\bf k}}\,$.
To this end, it is necessary to find the eigenvalues of the matrix (\ref{Eq3.8})
by solving the equation
\begin{equation}     \label{Eq3.10}
\det \, \left|\tilde{\mathfrak{M}}_{2}({\bf k}_{1},{\bf k}_{2};t)-
\tilde{E}({\bf k};t)\right|=0,
\end{equation}
$\tilde{E}({\bf k};t)$ is the diagonal matrix.
Further, the expression (\ref{Eq3.7}) can be written as follows:
\begin{align}    \label{Eq3.11}
W^{\text{G}}(a;t)&=\int\rd\tilde{\omega}\, \det \tilde{W}\,
 \exp \Bigg[ -\ri\pi \sum_{l,\bf k}   \tilde{a}_{l{\bf k}}
 \tilde{\omega}_{l{\bf k}}
- \, \frac{1}{2V^2}\sum_{\bf q}\sum_{\bf k}\beta_{-{\bf q}}(t)\nu({\bf k})
\left(n_{{\bf q}+{\bf k}} n_{-{\bf k}} - n_{{\bf q}}\right)                       \nonumber  \\
&\quad- \, \frac{\pi^{2}}{2}\sum_{l} \sum_{{\bf k}}
E_{l}({\bf k};t)\tilde{\omega}_{l{\bf k}} \tilde{\omega}_{l,-{\bf k}}\Bigg],
\end{align}
where new variables $\tilde{a}_{l{\bf k}}\,$, $\tilde{\omega}_{l{\bf k}}$
are connected with the old variables by the ratio:
\[\tilde{a}_{n{\bf k}}=\sum_{l} \bar{a}_{l{\bf k}}\omega_{ln} \,, \qquad
   \omega_{l{\bf k}}=\sum^{3}_{m=1}\omega_{lm}\tilde{\omega}_{m{\bf k}}\,,
\]
and $\omega_{lm}$ are matrix elements:
\begin{eqnarray}
\tilde{W}=\begin{vmatrix}
 \omega_{11}&  \ldots & \omega_{15} \\
  \vdots    &  \ddots & \vdots      \\
 \omega_{51} & \ldots & \omega_{55}
\end{vmatrix}_{({\bf k};t)}. \nonumber
\end{eqnarray}
Integrand in (\ref{Eq3.11}) is a quadratic function $\tilde{\omega}_{n{\bf k}}$ and
after integrating over $\rd\omega_{n{\bf k}}$ we will obtain the following structure function
in Gaussian approximation $W^{\text{G}}(a;t)$:
\begin{align}    \label{Eq3.12}
 W^{\text{G}}(a;t)&= \exp\Bigg[-\frac{1}{2}\sum_{l,\bf k} E^{-1}_{l}({\bf k};t)
  \tilde{a}_{l{\bf k}}\tilde{a}_{l,-{\bf k}}
- \frac{1}{2V^2}\sum_{\bf q}\sum_{\bf k}\beta_{-{\bf q}}(t)
  \nu({\bf k})\left(n_{{\bf q}+{\bf k}}n_{-{\bf k}} - n_{{\bf q}}\right) \Bigg]      \nonumber  \\
&\quad\times \exp\Bigg[-\frac{1}{2} \sum_{\bf k} \ln\,\pi \,\det \tilde{E}({\bf k};t) \Bigg]
  \exp \Bigg[\sum_{\bf k}\ln\,\det \tilde{W}({\bf k};t) \Bigg],
\end{align}
or through variables $\bar{a}_{l{\bf k}}$:
\begin{eqnarray}                   \label{Eq3.13}
 W^{\text{G}}(a;t)= Z(t)\,\exp \Bigg[ -\frac{1}{2}\sum_{l,\bf k} \bar{E}_{l}({\bf k};t)
  \bar{a}_{l{\bf k}}\bar{a}_{l,-{\bf k}}
-\frac{1}{2V^2}\sum_{\bf q}\sum_{\bf k}\beta_{-{\bf q}}(t)
  \nu({\bf k})\left(n_{{\bf q}+{\bf k}}n_{-{\bf k}} - n_{{\bf q}}\right) \Bigg],
\end{eqnarray}
where
\begin{equation}
\nonumber
\bar{E}_{l}({\bf k};t)=\sum_{l'}\omega_{ll'}E^{-1}_{l'}({\bf k};t)\omega_{l'l}\,,    \\
\end{equation}
\begin{equation}        \nonumber
Z(t)= \exp \Bigg[-\frac{1}{2} \sum_{\bf k} \ln\,\pi \,\det \tilde{E}({\bf k};t) \Bigg]
  \exp \Bigg[\sum_{\bf k} \ln\,\det \tilde{W}({\bf k};t) \Bigg].
\end{equation}
The structure function $W^{\text{G}}(a;t)$ provides a possibility to calculate
(\ref{Eq3.2}) in higher approximations over Gaussian moments
\cite{IdzIgnTok1996,HlushTokar2016}:
\begin{equation}    \label{Eq3.14}
 W(a;t)=W^{\text{G}}(a;t)\exp \left[\sum_{n\geqslant 3}\langle \tilde{D}_{n}(a ;t)\rangle_{\text{G}}\right],
\end{equation}
where one presents $\langle \tilde{D}_{n}(a ;t)\rangle_{\text{G}}$  approximately as:
\[
\langle\tilde{D}_{3}(a ;t)\rangle_{\text{G}}=\langle \bar{D}_{3}(a ;t)\rangle_{\text{G}}\,,
\]
\[
\langle\tilde{D}_{4}(a ;t)\rangle_{\text{G}}=\langle \bar{D}_{4}(a ;t)\rangle_{\text{G}}\,,
\]
\[
\langle\tilde{D}_{6}(a ;t)\rangle_{\text{G}}=\langle \bar{D}_{6}(a ;t)\rangle_{\text{G}}-
\frac{1}{2}\langle\bar{D}_{3}(a ;t)\rangle_{\text{G}}^{2}\,,
\]
\[
\langle\tilde{D}_{8}(a ;t)\rangle_{\text{G}}=\langle \bar{D}_{8}(a ;t)\rangle_{\text{G}}-
\langle\bar{D}_{3}(a;t)\rangle_{\text{G}}\langle \bar{D}_{5}(a ;t)\rangle_{\text{G}} -
\frac{1}{2}\langle \bar{D}_{4}(a ;t)\rangle_{\text{G}}^{2}\,,
\]
\[
\langle\tilde{D}_{n}(a;t)\rangle_{\text{G}}=
\frac{1}{W^{\text{G}}(a;t)}\sum_{l_{1},\ldots,l_{n}}\sum_{{\bf k}_{1},\ldots,{\bf k}_{n}}
\bar{\mathfrak{M}}_{n}^{l_{1},\ldots,l_{n}}({\bf k}_{1},\ldots,{\bf k}_{n};t)
\frac{1}{(\ri\pi)^{n}}\frac{\delta^{n}}{\delta \bar{a}_{l_{1},{\bf k}_{1}}\ldots
\delta \bar{a}_{l_{n},{\bf k}_{n}}}W^{\text{G}}(a;t).
\]

\noindent
$\langle\tilde{D}_{n}(a;t)\rangle_{\text{G}}$ are the renormalized
non-equilibrium cumulant averages of the order $n$ for
the variables $\bar{a}_ {l{\bf k}}\,$.
In expression (\ref{Eq3.14}), the summands
are with only even degrees over ${a}$ since all odd Gaussian moments vanish.

The method of calculation of the structure function $W(a;t)$ can be used for approximate
calculations of hydrodynamic velocities $v_{l,{\bf k}}(a;t)$. We present a general formula
of velocities consistent with (\ref{Eq2.32}) in the following form:
\begin{equation}        \label{Eq3.15}
v_{l,{\bf k}}(a;t)=\int \rd\Gamma_{N}\dot{\hat{a}}_{l,{\bf k}}
    \varrho_{\text{rel}}^{\text{kin-hyd}} \big(x^{N};t\big)\hat{f}(a)
\end{equation}
and introduce the function $W(a,\lambda;t)$:
\begin{equation}        \label{Eq3.16}
W(a,\lambda;t)=\int\rd\Gamma_{N}\,\,\exp\Bigg(-\ri\pi\sum_{l,\bf k}\lambda_{l,\bf k}\dot{\hat{a}}_{l,{\bf k}}\Bigg)\,\,\varrho_{\text{rel}}^{\text{kin-hyd}} \big(x^{N};t\big)\hat{f}(a),
\end{equation}
so that
\begin{equation}    \label{Eq3.17}
v_{l,{\bf k}}(a;t)=\frac{\partial }{\partial (-\ri\pi\lambda_{l,\bf k})}
\ln W(a,\lambda;t)\Big|_{\lambda_{l,\bf k}=0}.
\end{equation}
We calculate the function $W(a,\lambda;t)$ using the preliminary results
of the calculation of the structural function $W(a;t)$,
and rewrite $W(a,\lambda;t)$ as:
\begin{eqnarray}    \label{Eq3.18}
 W(a,\lambda;t)=\int\rd\Gamma_{N}\int\rd\omega
 \exp \Bigg(-\ri\pi\sum_{l,\bf k}\lambda_{l,\bf k}\dot{\hat{a}}_{l,{\bf k}}\Bigg)
\,\exp \Bigg[ -\ri\pi \sum_{l,\bf k}\omega_{l,\bf k}\left(\hat{a}_{l,\bf k}-a_{l,\bf k}\right)\Bigg]
 \varrho_{\text{rel}}^{\text{kin-hyd}} \big(x^{N};t\big).
\end{eqnarray}
Now, we carry out an averaging in (\ref{Eq3.18}) using the following cumulant expansion:
\begin{align}    \label{Eq3.19}
W(a,\lambda;t)&=
\int \rd\omega \exp \Bigg\{-\ri\pi \sum_{l,\bf k}\omega_{l,\bf k}\bar{a}_{l,\bf k}
- \frac{1}{2V^2}\sum_{\bf q}\sum_{\bf k}\beta_{-{\bf q}}(t)\nu({\bf k})\left(n_{{\bf q}+
  {\bf k}}n_{-{\bf k}} - n_{{\bf q}}\right)   \nonumber \\
&\quad+ \sum_{n\geqslant 1}\big[D_{n}(\omega ;t)+D_{n}(\lambda ;t)+D_{n}(\omega,\lambda ;t)\big]\Bigg\},
\end{align}
where
\begin{align}         \nonumber
 D_{n}(\omega ;t)&=\frac{(-\ri\pi)^{n}}{n!}\sum_{l_{1},\ldots,l_{n}}\sum_{{\bf k}_{1},\ldots,
   {\bf{k}}_{n}}   \mathfrak{M}_{n}^{l_{1},\ldots,l_{n}}({\bf k}_{1},\ldots,{\bf k}_{n};t)
   \omega_{l_{1},{\bf k}_{1}}\ldots \omega_{l_{n},{\bf k}_{n}},       \\
\nonumber
D_{n}(\lambda ;t)&=\frac{(-\ri\pi)^{n}}{n!}\sum_{l_{1},\ldots,l_{n}}\sum_{{\bf k}_{1},\ldots,
{\bf k}_{n}} \mathfrak{M}_{n}^{(1)l_{1},\ldots,l_{n}}({\bf k}_{1},\ldots,
{\bf k}_{n};t)\lambda_{l_{1},{\bf k}_{1}}\ldots\lambda_{l_{n},{\bf k}_{n}},     \\
\nonumber
D_{n}(\omega, \lambda ;t)&=\frac{(-\ri\pi)^{n}}{n!}
\sum_{l_{1},\ldots,l_{n}}\sum_{{\bf k}_{1},\ldots,{\bf k}_{n}}
 \mathfrak{M}_{n}^{(2)l_{1},\ldots,l_{n}}({\bf k}_{1},\ldots,{\bf k}_{n};t) \omega_{l_{1},{\bf k}_{1}}\ldots \omega_{l_{n-1},{\bf k}_{n-1}}
  \ldots\lambda_{l_{n},{\bf k}_{n}},
\end{align}
with the cumulants of the following structure:
\begin{align}
\nonumber
&\mathfrak{M}_{n}^{l_{1},\ldots,l_{n}}({\bf k}_{1},\ldots,{\bf k}_{n};t)=
\langle \hat{a}_{l_{1},{\bf k}_{1}},\ldots,\hat{a}_{l_{n},{\bf k}_{n}}\rangle_{\text{kin-sh}}^{t,\text{c}},  \\
\nonumber
& \mathfrak{M}_{n}^{(1)l_{1},\ldots,l_{n}}({\bf k}_{1},\ldots,{\bf k}_{n};t)=
 \langle\dot{\hat{a}}_{l_{1},{\bf k}_{1}},\ldots,\dot{\hat{a}}_{l_{n},{\bf k}_{n}}\rangle_{\text{kin-sh}}^{t,\text{c}},                                                   \\
\nonumber
 &\mathfrak{M}_{n}^{(2)l_{1},\ldots,l_{n}}({\bf k}_{1},\ldots,{\bf k}_{n};t)=n\big[(n-j)+(j-n+1)
 \delta_{l_{1},\ldots,l_{n-1}}\big]                    \\
\nonumber
&\times\langle \hat{a}_{l_{1},{\bf k}_{1}},\ldots,\hat{a}_{l_{n-j},{\bf k}_{n-j}},\ldots,
 \dot{\hat{a}}_{l_{n-j+1},{\bf k}_{n-j+1}},\ldots,\dot{\hat{a}}_{l_{n},{\bf k}_{n}}\rangle_{\text{kin-sh}}^{t,\text{c}}.
\end{align}

First, we consider a Gaussian approximation for $W(a,\lambda;t)$,
namely in the exponent of
an integrand we leave only the summands with $n=2$ and linear over $\lambda_{l,{\bf k}}$:
\begin{align}    \label{Eq3.20}
  W^{\text{G}}(a,\lambda;t)&=\int\rd\omega\exp\Bigg[ \ri\pi \sum_{l,{\bf k}}
  \omega_{l,{\bf k}}\bar{a}_{l,\bf k}-\ri\pi \sum_{l,\bf k}
  \langle\dot{\hat{a}}_{l,{\bf k}}\rangle_{\text{kin-hyd}}^{t,\text{c}}\lambda_{l,{\bf k}}
-\frac{1}{2V^2}\sum_{\bf q}\sum_{\bf k}\beta_{-{\bf q}}(t)\nu({\bf k})
  \left(n_{{\bf q}+{\bf k}}n_{-{\bf k}} - n_{{\bf q}}\right)  \nonumber \\
&\quad-   \frac{\pi^{2}}{2}\sum_{l_{1},l_{2}}\sum_{{\bf k}_{1},{\bf k}_{2}}
  \mathfrak{M}_{2}^{l_{1},l_{2}}({\bf k}_{1},{\bf k}_{2};t)
  \omega_{l_{1},{\bf k}_{1}} \omega_{l_{2},{\bf k}_{2}}
- \frac{\pi^{2}}{2}\sum_{l_{1},l_{2}}\sum_{{\bf k}_{1},{\bf k}_{2}}
  \mathfrak{M}_{2}^{(2)l_{1},l_{2}}({\bf k}_{1},{\bf k}_{2};t)
  \omega_{l_{1},{\bf k}_{1}} \lambda_{l_{2},{\bf k}_{2}}\Bigg].
\end{align}
Then, transforming this expression in the exponent to a diagonal quadratic form over variables
$\omega_{l,{\bf k}}\,$, similarly to $W(a;t)$, after integrating with respect to the new variables $\bar{\omega}_{l,{\bf k}}\,$, one obtains:
\begin{align}    \label{Eq3.21}
 W^{\text{G}}(a,\lambda;t)&=\int\rd\omega \exp \Bigg[-\ri\pi \sum_{l,{\bf k}}
  \langle\dot{\hat{a}}_{l,{\bf k}}\rangle_{\text{kin}}^{t}\lambda_{l,{\bf k}}
- \frac{1}{2V^2}\sum_{\bf q}\sum_{\bf k}\beta_{-{\bf q}}(t)\nu({\bf k})
     \left(n_{{\bf q}+{\bf k}}n_{-{\bf k}} - n_{{\bf q}}\right)   \nonumber  \\
&\quad- \frac{\pi^{2}}{2}\sum_{l,{\bf k}}E_{l}^{-1}({\bf k};t)b_{l,{\bf k}}b_{l,-{\bf k}}
- \frac{1}{2}\sum_{{\bf k}}\ln\pi \det\tilde{E}({\bf k};t)+\sum_{{\bf k}}
\ln \det\tilde{W}({\bf k};t)\Bigg],
\end{align}
where
\[
b_{l,{\bf k}}=\sum_{j}\omega_{lj}\Bigg[\bar{a}_{j,{\bf k}}+
\frac{\ri\pi}{2}\sum_{j'}\mathfrak{M}_{2}^{(2)j,j'}({\bf k};t)\lambda_{j',{\bf k}}\Bigg],
\]
and $\omega_{lj}$, $\mathfrak{M}_{2}^{(2)j,j'}({\bf k};t)$ and $E_{l}({\bf k};t)$ are
not dependent on $\lambda_{l,{\bf k}}\,$.
Here, the cumulants  $\mathfrak{M}_{2}^{(2)j,j'}({\bf k};t)$ have the following structure:
\begin{equation}    \label{Eq3.22}
\mathfrak{M}_{2}^{(2)j,j'}({\bf k};t)=
    \langle\dot{\hat{a}}_{j,{\bf k}}\hat{a}_{j',-{\bf k}}\rangle_{\text{kin-sh}}^{t}-
    \langle\dot{\hat{a}}_{j,{\bf k}}\rangle_{\text{kin}}^{t}\langle\hat{a}_{j',-{\bf k}}\rangle_{\text{kin-sh}}^{t}.
\end{equation}
Now, we calculate the hydrodynamic velocities $v_{l,{\bf k}}(a;t)$ in a Gaussian approximation
according to the formula
\begin{eqnarray}    \label{Eq3.23}
v_{l,{\bf k}}(a;t)=\frac{\partial }{\partial (-\ri\pi\lambda_{l,\bf k})}
\ln W^{\text{G}}(a,\lambda;t)\Big|_{\lambda_{l,{\bf k}}=0}
=\,\langle\dot{\hat{a}}_{j,{\bf k}}\rangle_{\text{kin}}^{t}-
\frac{1}{2}\sum_{j,j'}E_{l}^{-1}({\bf k};t)\omega_{jl}
\omega_{j'l}\mathfrak{M}_{2}^{(2)j',l}({\bf k};t)\bar{a}_{l,{\bf k}}\,.
\end{eqnarray}
Specifically, we consider the particular case when one can divide the longitudinal
and transverse fluctuations for collective variables.
That is, we choose the direction of the wave vector ${\bf k}$ along the axis of $Oz$.
Thus, one obtains:
\begin{align}    \label{Eq3.24}
W^{\text{G}}(a;t)&=\int\rd\omega \exp\Bigg[\ri\pi \sum_{l,\bf k}\omega_{l,\bf k}\bar{a}_{l,\bf k}
-   \frac{1}{2V^2}\sum_{\bf q}\sum_{\bf k}\beta_{-{\bf q}}(t)\nu({\bf k})\left(n_{{\bf q}+{\bf k}}n_{-{\bf k}} -
  n_{{\bf q}}\right) \nonumber \\
&\quad- \frac{\pi^{2}}{2}\sum_{l_{1},l_{2}=1}^{3}\sum_{{\bf k}_{1},{\bf k}_{2}}
  \mathfrak{M}^{\parallel, l_{1},l_{2}}_{2}({\bf k}_{1},{\bf k}_{2};t)
  \omega_{l_{1},{\bf k}_{1}} \omega_{l_{2},{\bf k}_{2}}      \nonumber \\
&\quad
- \frac{\pi^{2}}{2}\sum_{l_{1},l_{2}=1}^{4}\sum_{{\bf k}_{1},{\bf k}_{2}}
  \mathfrak{M}^{\parallel,\perp, l_{1},l_{2}}_{2}({\bf k}_{1},{\bf k}_{2};t)
  \omega_{l_{1},{\bf k}_{1}} \omega_{l_{2},{\bf k}_{2}}\Bigg],
\end{align}
where $\mathfrak{M}^{\parallel, l_{1},l_{2}}_{2}({\bf k}_{1},{\bf k}_{2};t)$
are the matrix elements of the non-equilibrium correlation functions
of longitudinal fluctuations
\begin{equation}    \label{Eq3.25}
\mathfrak{M}^{\parallel}_{2}({\bf k}_{1},{\bf k}_{2};t)=
\begin{vmatrix}
 \langle \hat{n}\hat{n}\rangle_{\text{kin-sh}}^{\text{c}}&
 \langle \hat{n}\hat{{\bf J}}^{\parallel}\rangle_{\text{kin-sh}}^{\text{c}} &
 \langle \hat{n}\hat{\varepsilon}\rangle_{\text{kin-sh}}^{\text{c}}  \vspace{1mm} \\
  \langle \hat{{\bf J}}^{\parallel}\hat{n}\rangle_{\text{kin-sh}}^{\text{c}}        &
  \langle \hat{{\bf J}}^{\parallel}\hat{{\bf J}}^{\parallel}\rangle_{\text{kin-sh}}^{\text{c}}  &
  \langle \hat{{\bf J}}^{\parallel}\hat{\varepsilon}\rangle_{\text{kin-sh}}^{\text{c}} \vspace{1mm}\\
  \langle \hat{\varepsilon}\hat{n}\rangle_{\text{kin-sh}}^{\text{c}}  &
  \langle \hat{\varepsilon}\hat{{\bf J}}^{\parallel}\rangle_{\text{kin-sh}}^{\text{c}}  &
  \langle \hat{\varepsilon}\hat{\varepsilon}\rangle_{\text{kin-sh}}^{\text{c}}
  \end{vmatrix}_{{\bf k_1,k_2}},
\end{equation}
$\mathfrak{M}^{\perp l_{1},l_{2}}_{2}({\bf k}_{1},{\bf k}_{2};t)$
are the matrix elements of the non-equilibrium correlation functions of transverse and
transverse-longitudinal fluctuations
\begin{eqnarray}    \label{Eq3.26}
\mathfrak{M}^{\parallel,\perp}_{2}({\bf k}_{1},{\bf k}_{2};t)=
\begin{vmatrix}
 0&
 \langle \hat{n}\hat{{\bf J}}^{\perp}_{x}\rangle_{\text{kin-sh}}^{\text{c}} &
 \langle \hat{n}\hat{{\bf J}}^{\perp}_{y}\rangle_{\text{kin-sh}}^{\text{c}} &
 0   \vspace{1mm} \\
 \langle \hat{{\bf J}}^{\perp}_{x}\hat{n}\rangle_{\text{kin-sh}}^{\text{c}}        &
  \langle \hat{{\bf J}}^{\perp}_{x}\hat{{\bf J}}^{\perp}_{x}\rangle_{\text{kin-sh}}^{\text{c}}  &
  \langle \hat{{\bf J}}^{\perp}_{x}\hat{{\bf J}}^{\perp}_{y}\rangle_{\text{kin-sh}}^{\text{c}}  &
  \langle \hat{{\bf J}}^{\perp}_{x}\hat{\varepsilon}\rangle_{\text{kin-sh}}^{\text{c}}  \vspace{1mm} \\
  \langle \hat{{\bf J}}^{\perp}_{y}\hat{n}\rangle_{\text{kin-sh}}^{\text{c}}        &
  \langle \hat{{\bf J}}^{\perp}_{y}\hat{{\bf J}}^{\perp}_{x}\rangle_{\text{kin-sh}}^{\text{c}}  &
  \langle \hat{{\bf J}}^{\perp}_{y}\hat{{\bf J}}^{\perp}_{y}\rangle_{\text{kin-sh}}^{\text{c}}  &
  \langle \hat{{\bf J}}^{\perp}_{y}\hat{\varepsilon}\rangle_{\text{kin-sh}}^{\text{c}} \vspace{1mm}  \\
  0  &
  \langle \hat{\varepsilon}\hat{{\bf J}}^{\perp}_{x}\rangle_{\text{kin-sh}}^{\text{c}}  &
  \langle \hat{\varepsilon}\hat{{\bf J}}^{\perp}_{y}\rangle_{\text{kin-sh}}^{\text{c}}  &
  0
\end{vmatrix}_{{\bf k_1,k_2}}.
\end{eqnarray}
In this case, the hydrodynamic velocities in the Gaussian approximation are as follows:
\begin{align}    \label{Eq3.27}
v_{n{\bf k}}^{\parallel \text{G}}(a;t)&=\langle\dot{\hat{n}}_{{\bf k}}\rangle_{\text{kin-sh}}^{t}+ E_{1}^{-1}({\bf
k};t)\big(\omega_{11}\bar{n}_{{\bf k}}+ \omega_{21}\bar{{\bf J}}^{\parallel}_{{\bf k}} + \omega_{31}
\bar{\varepsilon}_{{\bf k}}\big)\Omega_{n}({\bf k};t),    \\
\nonumber
v_{J{\bf k}}^{\parallel \text{G}}(a;t)&=\langle\dot{\hat{{\bf J}}}^{\parallel }_{{\bf k}}\rangle_{\text{kin-sh}}^{t}+
E_{2}^{-1}({\bf k};t)\big(\omega_{12}\bar{n}_{{\bf k}}+\omega_{22}\bar{{\bf J}}^{\parallel }_{{\bf k}} +
\omega_{32} \bar{\varepsilon}_{{\bf k}}\big)\Omega_{J}({\bf k};t),  \\
\nonumber
v_{\varepsilon{\bf k}}^{\parallel \text{G}}(a;t)&=\langle\dot{\hat{\varepsilon}}_{{\bf k}}\rangle_{\text{kin-sh}}^{t}+
E_{3}^{-1}({\bf k};t)\big(\omega_{13}\bar{n}_{{\bf k}}+ \omega_{23}\bar{{\bf J}}^{\parallel}_{{\bf k}} +
\omega_{33} \bar{\varepsilon}_{{\bf k}}\big)\Omega_{\varepsilon}({\bf k};t),
\end{align}
where
\begin{align}           \label{Eq3.28}
\Omega_{n}({\bf k};t)&=\omega_{11}\langle \hat{n}_{{\bf k}}\dot{\hat{n}}_{-{\bf k}}
\rangle_{\text{kin-sh}}^{t,\text{c}}+ \omega_{21}\langle \hat{{\bf J}}^{\parallel}_{{\bf k}}
\dot{\hat{n}}_{-{\bf k}}
\rangle_{\text{kin-sh}}^{t,\text{c}}+ \omega_{31}\langle \hat{\varepsilon}_{{\bf k}}
   \dot{\hat{n}}_{-{\bf k}}
\rangle_{\text{kin-sh}}^{t,\text{c}},                    \\
\nonumber
\Omega_{J}({\bf k};t)&=\omega_{12} \langle \hat{n}_{{\bf k}}
\dot{\hat{{\bf J}}}^{\parallel}_{-{\bf k}}
\rangle_{\text{kin-sh}}^{t,\text{c}}+\omega_{22}\langle\hat{{\bf J}}^{\parallel}_{{\bf k}}
\dot{\hat{{\bf J}}}^{\parallel}_{-{\bf k}}
\rangle_{\text{kin-sh}}^{t,\text{c}}+\omega_{32}\langle\hat{\varepsilon}_{{\bf k}}
\dot{\hat{{\bf J}}}^{\parallel}_{-{\bf k}} \rangle_{\text{kin-sh}}^{t,\text{c}},                  \\
\nonumber
 \Omega_{\varepsilon}({\bf k};t)&=\omega_{13} \langle \hat{n}_{{\bf k}}
 \dot{\hat{\varepsilon}}_{-{\bf k}}\rangle_{\text{kin-sh}}^{t,\text{c}}+
 \omega_{23} \langle \hat{{\bf J}}^{\parallel}_{{\bf k}}
 \dot{\hat{\varepsilon}}_{-{\bf k}}\rangle_{\text{kin-sh}}^{t,\text{c}}+
 \omega_{33}\langle\hat{\varepsilon}_{{\bf k}}
 \dot{\hat{\varepsilon}}_{-{\bf k}} \rangle_{\text{kin-sh}}^{t,\text{c}},
\end{align}
and $\omega_{lj}$ are the elements of matrix $\tilde{W}({\bf k};t)$.

As one can see the hydrodynamic velocities (\ref{Eq3.27}) in the Gaussian approximation for
$W^{\text{G}}(a,\lambda;t)$ are linear functions of the collective variables $n_{\bf k}\,$,
${\bf J}_{\bf k}$ and $\varepsilon_{\bf k}$.
It is remarkable that if the kinetic processes are not taken into account, i.e.,
$\varrho_{\text{rel}}^{\text{kin-sh}} (x^{N};t)=1$, then
$\langle\ldots\rangle_{\text{kin-sh}}^{t}\rightarrow\langle\ldots\rangle_{0}$
is an average over a microscopic ensemble $W(a)$;
in this case, the expressions (\ref{Eq3.27}) for hydrodynamic velocities
transform into the results of the previous work \cite{IdzIgnTok1996},
in which the nonlinear hydrodynamic fluctuations in simple fluids were investigated.

\section{Conclusions}

Using the method of Zubarev non-equilibrium statistical operator,
we have obtained a modified chain of BBGKY kinetic equations that take into account
non-equilibrium hydrodynamic fluctuations for a system of interacting particles.
At the same time, the non-equilibrium distribution function that describes
the hydrodynamic fluctuations, satisfies a generalized Fokker-Planck equation.
We divide the contributions from short-range and long-range interactions between particles,
and describe the short-range interactions (hard sphere model) in the coordinate space,
while the long-range interactions are described in the space of collective variables.

Moreover, the short-range component will be considered as a basis with distribution
$\varrho_{\text{rel}}^{\text{kin-sh}} (x^{N};t)$, which corresponds
to the BBGKY chain of equations for the model of hard spheres \cite{KobrOmelTok1998}.

The used method of collective variables \cite{Zubarev1982,YukhnHolov1980,HlushTokar2016}
has made it possible to calculate
the structural distribution function of hydrodynamic collective variables and their
hydrodynamic velocities in approximations higher than the Gaussian one.
In particular, the hydrodynamic velocities above the Gaussian approximation that follow
from equation~(\ref{Eq3.19}) and (\ref{Eq3.27}) will be proportional to
$\Bar{a}_{l,{\bf k}}\bar{a}_{l',{\bf k}} \,$, and the transport kernels in the
Fokker-Planck equation will be of fourth-order correlation functions
over variables $\hat{a}_{l,{\bf k}}\,$.
It is significant that the Fokker-Planck equation in a Gauss approximation for
$\tilde{W}^{\text{G}}({\bf k};t)$, $v_{l,{\bf k}}^{\text{G}}(a;t)$ leads to transport equations
for $\langle \hat{a}_{l,{\bf k}}\rangle^{t}$ and  the structure of these equations is
the same as in molecular hydrodynamics \cite{MrygTok1992,MrygOmelTok1995},
although the averaging  is performed  using
$\varrho_{\text{L}}(x^{N},a;t)=\varrho_{\text{rel}}^{\text{kin-hyd}}(x^{N};t)\hat{f}(a)/ W^{\text{G}}(a;t)$.

The proposed approach makes it possible to go beyond the Gaussian approximation for
$\tilde{W}({\bf k};t)$, $v_{l,{\bf k}}(a;t)$ and for transport kernels
in the Fokker-Planck equation.
As a result, we obtain a nonlinear system of equations  for the
$\langle \hat{a}_{l,{\bf k}}\rangle^{t}$.
It is remarkable that the kinetic equation (\ref{Eq2.41}) contains generalized integrals of
Fokker-Planck type with generalized coefficients of diffusion and friction
in the phase space $({\bf r},{\bf p},t)$, where the limit of change of $|{\bf r}|$
is restricted by value $|{\bf k}|^{-1}_{\text{hydr}}$ that corresponds to collective
non-linear hydrodynamic fluctuations.
This means that in areas smaller than $|{\bf k}|^{-1}_{\text{hydr}}\,$, the processes are described
by generalized coefficients of diffusion and friction, while in areas larger than
$|{\bf k}|^{-1}_{\text{hydr}}$ they are described by generalized coefficient
of viscosity, thermal conductivity $\phi_{j j}({\bf k},{\bf q},a,a';t,t')$,
$\phi_{\varepsilon \varepsilon}({\bf k},{\bf q},a,a';t,t')$
and by cross coefficients
$\phi_{j \varepsilon}({\bf k},{\bf q},a,a';t,t')$,
$\phi_{ \varepsilon j}({\bf k},{\bf q},a,a';t,t')$.
The correlations between these areas of transport are described by kernels
$\phi_{n j}(x,{\bf q},a,a';t,t')$, $\phi_{n \varepsilon}(x,{\bf q},a,a';t,t')$,
$\phi_{ \varepsilon n}({\bf k},x',a,a';t,t')$, \linebreak $\phi_{jn}({\bf k},x',a,a';t,t')$,
which are present both in kinetic and Fokker-Planck equations, and
describe cross-correlation between kinetic and hydrodynamic processes.

\ukrainianpart

\title{Ланцюжок кінетичних рівнянь ББГКІ, метод нерівноважного статистичного оператора та метод
колективних змінних в нерівноважній статистичній теорії рідин}

\author{І.Р.~Юхновський, П.А.~Глушак, М.В.~Токарчук }
\address{Інститут фізики конденсованих систем НАН України,
вул. Свєнціцького, 1, 79011 Львів, Україна}
\makeukrtitle

\begin{abstract}
Запропоновано ланцюжок кінетичних рівнянь для нерівноважних одночастинкової,
двочастинкової і $s$-частинкової функцій розподілу частинок з урахуванням нелінійних
гідродинамічних флуктуацій.
Використовується метод нерівноважного статистичного оператора Зубарєва з проектуванням.
Нелінійні гідродинамічні флуктуації описуються нерівноважною функцією розподілу колективних
змінних, що задовольняє узагальнене рівняння Фоккера-Планка.
На основі методу колективних змінних запропоновано спосіб розрахунку нерівноважної структурної
функції розподілу  колективних змінних та їх гідродинамічних швидкостей (вище гаусового
наближення), що містяться в узагальненому рівнянні Фоккера-Планка для нерівноважної функції
розподілу  колективних змінних.
При цьому  розділені  вклади від короткодіючих і далекодіючих взаємодій між частинками,
що привело до того, що короткодіючі взаємодії (наприклад, модель твердих сфер)
описуються в координатному просторі, а далекодіючі --- у просторі колективних змінних.
Короткодіюча складова  розглядається як базисна, якій відповідає ланцюжок рівнянь ББГКІ
для моделі твердих сфер.
\keywords нелінійні флуктуації, нерівноважний статистичний оператор, функція розподілу,
          рівняння Фоккера-Планка, проста рідина

\end{abstract}
\end{document}